\documentclass[usenatbib]{mn2e}
\usepackage{astrojournals}
\usepackage{graphicx}
\usepackage{multirow}
\usepackage{hyperref}
\usepackage{url} 
\usepackage{amsmath}
\usepackage{times}
\usepackage{amssymb}
\usepackage{booktabs}
\usepackage{tabularx}
\usepackage{array}

\def\gvol{\mathcal{V}}
\def\chvol{\mathcal{V}_{CH}}
\def\grvol{\mathcal{V}_{MC}}
\def\evol{\mathcal{V}_{ME}}
\def\lvol{\mathcal{V}}
\def\ineff{\eta}
\def\npart{N_{p}}
\def\zini{z_{ini}}
\def\mvir{\mathrm{M}_{\mathrm{v}}} 
\def\rvir{\mathrm{R}_{\mathrm{v}}} 
\def\vvir{\mathrm{V}_{\mathrm{v}}} 
\def\volvir{(4\pi\rvir^{3}/3)} 
\def\rtb{\mathrm{R}_{\mathrm{tb}}} 
\def\enzo{\texttt{ENZO}}
\def\gadget{\texttt{GADGET-2}}
\def\music{\texttt{MUSIC}}
\def\hmpc{h^{-1}\mathrm{Mpc}}
\def\hmsun{h^{-1} M_{\sun}}
\newcommand{\lagvol}{Lagrange volume}

\title[How to Zoom]{How to Zoom: Bias, Contamination, and Lagrange Volumes in Multimass Cosmological Simulations}
\author[J.~O\~norbe et al.]{
Jose~O\~norbe,$^{1,2}$\thanks{E-mail: \texttt{jonorbeb@uci.edu}}
Shea~Garrison-Kimmel,$^1$
Ariyeh~H.~Maller,$^{3,4}$\newauthor
James~S.~Bullock,$^{1,2}$ 
Miguel~Rocha,$^1$ and
Oliver~Hahn$^5$\\
$^1$Center for Cosmology, Department of Physics and Astronomy, The University of California at Irvine, Irvine, CA, 92697, USA\\
$^2$Center for Galaxy Evolution, Department of Physics and Astronomy, The University of California at Irvine, Irvine, CA, 92697, USA\\
$^3$Department of Physics, New York City College of Technology, CUNY, Brooklyn, NY 11201, USA\\
$^4$Department of Astrophysics, American Museum of Natural History, New York, NY 10024, USA\\
$^5$ETH Zurich Institute for Astronomy. CH-8093 Zurich. Switzerland}

\begin{document}
\label{firstpage}

\maketitle

\begin{abstract}{We perform a suite of multimass cosmological zoom simulations of individual dark matter halos and explore how to best select Lagrangian regions for resimulation without contaminating the halo of interest with low-resolution particles. Such contamination can lead to significant errors in the gas distribution of hydrodynamical simulations, as we show. For a fixed \lagvol{}, we find that the chance of contamination increases systematically with the level of zoom. In order to avoid contamination, the Lagrangian volume selected for resimulation must increase monotonically with the resolution difference between parent box and the zoom region.  We provide a simple formula for selecting Lagrangian regions (in units of the halo virial volume) as a function of the level of zoom required. We also explore the degree to which a halo's Lagrangian volume correlates with other halo properties (concentration, spin, formation time, shape, etc.) and find no significant correlation. There is a mild correlation between \lagvol{} and environment, such that halos living in the most clustered regions have larger Lagrangian volumes.  Nevertheless, selecting halos to be isolated is not the best way to ensure inexpensive zoom simulations.   We explain how one can safely choose halos with the smallest Lagrangian volumes, which are the least expensive to resimulate, without biasing one's sample.}
\end{abstract}

\begin{keywords}
cosmology: theory --- galaxies: formation --- galaxies: halos --- methods: $N$-body simulations  
\end{keywords}

\section{Introduction}
\label{sec:Introduction}

Over the past three decades, numerical simulations of cosmic structure formation have become a standard tool for studying a wide range of problems in cosmology.  The predictive power of such simulations increases with their complexity, but even with the increasing size and computational power of the machines used to run them, advances in computational algorithms have played a key role in pushing the limits of this method. For example, the multimass technique \citep{Porter:1985,Katz:1993,Navarro:1994} has converged as a common approach to study the role of physical processes important to galaxy formation in a cosmological context. 

The basic method behind this ``zoom" technique involves selecting a specific interesting region from a low resolution simulation and re-running it with higher resolution particles in that region. In this way, computational resources are focused mainly on the area of interest while the long range forces of gravity are still captured in their appropriate cosmological context
\citep{Navarro:1994,Frenk:1996,Tormen:1997,Thacker:2000,Bertschinger:2001,Klypin:2001,Power:2003,Springel:2008,Jenkins:2010,Hahn:2011}.
Currently, this technique is widely used to study the formation and evolution of all types of objects in a cosmological context, from dwarf galaxy halos to large clusters.

The region of interest is usually the virialized region of a dark matter halo, though, depending on the purpose of the simulation, it can be larger \citep[e.g.,][]{Hahn:2010,Onorbe:2011,GarrisonKimmel:2013}. Once the region of interest is selected, particles in that volume are traced back to their initial positions, thus defining a \emph{\lagvol{}} (so called because its key feature is that the particles trace the flow of matter through cosmic time). One then creates initial conditions with higher spatial and mass resolution in the \lagvol{}, while elsewhere retaining the lower resolution of the original box, such that long range forces from surrounding matter are properly resolved at minimal CPU and memory cost. In addition, several boundary or buffer regions are typically initialized at intermediate resolution around the \lagvol{}.  This method allows one to use a periodic box large enough to obtain cosmological convergence, including low frequency modes \citep{Power:2006}, while simultaneously reaching resolutions high enough to study the small scale processes relevant to galaxy formation.

Generating initial conditions for such a simulation requires the following steps:

\begin{description}
\item[\textbf{i:}] Determine the box size and resolution required in both the full-box and multimass simulations to study the given problem.
\item[\textbf{ii:}] Select a cosmological model ($\Omega_M$, $\Omega_\Lambda$, $\sigma_8$, etc.), calculate the appropriate initial redshift, and create a matter power spectrum (or transfer function) for those parameters at that redshift to generate the initial density fluctuations.
\item[\textbf{iii:}] Identify the region of interest in which higher resolution is desired. Usually this region is selected from a lower resolution full-box simulation by selecting a certain volume around a galactic halo. Particles inside this volume are identified and traced back to the initial conditions, and the \lagvol{} is then defined by the initial positions of those particles.
\item[\textbf{iv:}] Discretize the mass distribution in the simulation, according to the desired resolution, with the appropriate density fluctuations generated using the initial redshift matter power spectrum. 
\end{description}

Motivated primarily by the concern that the original approaches might be too simplistic and not accurate enough for the precision needed in full-box and multimass cosmological simulations, the preceding steps have been extensively refined over the last several years.  Several authors have contributed to this effort to a varying degree of detail; see, e.g. \citet{Navarro:1994,Frenk:1996,Tormen:1997,Thacker:2000,Bertschinger:2001,Klypin:2001,Power:2003,Springel:2008,Jenkins:2010,Hahn:2011}.

In particular, theorists have put a great deal of effort into improving the final step listed above which applies to both full-box and multimass simulations. These efforts fall into two categories: first, simulators have worked to properly generate the perturbations of the distribution from the overdensity field using second order perturbation theory \citep{Sirko:2005,Joyce:2009,Jenkins:2010} 
and second, they have improved the numerical techniques used to set the discrete distribution of mass, in particular by covolving the transfer function with (typically Gaussian) white noise to generate the correct overdensity field \citep{Bertschinger:2001,Hahn:2011}.

However, many of the choices that one must make in order to create multimass initial conditions have not been discussed before in the literature.  For example:  How large does the \lagvol{} need to be in order to avoid contamination in the high-resolution region by low-resolution particles?  Can one chose halos with intrinsically small \lagvol{}s in order to save cost without biasing results?  At what redshift must one start the resimulation?   How many buffer regions are necessary and how large should they be?   Most groups who perform multimass simulations have arrived at their methodologies by trial and error and/or the wisdom of experience.

In this paper, we improve upon this situation by testing the effects of these choices and quantifying their relative importance.  The paper is organized as follows.  In Section~\ref{sec:sims} we discuss the simulations we have run and the tools we have used to analyze them.  In Section~\ref{sec:identifying} we discuss methods for identifying the Lagrangian region of interest.  In Section~\ref{sec:contamination} we discuss how to avoid low-resolution particle contamination in the high-resolution region.  Section~\ref{sec:lagvol} shows that the \lagvol{} of a halo does not correlate with other halo properties except that objects in high-density environments tend to have higher \lagvol{}.  The implication is that halos with small \lagvol{} regions can be chosen for resimulation without biasing the outcome. 
Section~\ref{sec:createICs} discusses how the initial redshift of the zoom simulation and buffer regions around the \lagvol{} should be chosen.  We close with conclusions in Section~\ref{sec:conc}.

\section{Simulations}
\label{sec:sims}

We have run a series of cosmological N-body simulations following the formation and evolution of structure in the $\Lambda$CDM model. We use a sequence of boxes of side length $L_{box}$ that vary between 5 and 900 $\hmpc$. In each case we assume a flat cosmology with a cosmological constant term based on the latest results of WMAP \citep{Komatsu:2011}, with cosmological parameters\footnote{More details on the specific values considered can found at this \href{http://lambda.gsfc.nasa.gov/product/map/dr4/params/lcdm\_sz\_lens\_wmap7.cfm}{link}.}
$\Omega_{\Lambda}=0.734$, $\Omega_{m}=0.266$, $\Omega_{b}=0.0449$, $h=0.71$, normalization of an initial power-law index $n=0.963$ and $\sigma_{8}=0.801$. 

\begin{table}
\caption{Full-box simulations. First column: Name of the simulation. Second column: Size of the simulation box. Third column: Number of particles used in the simulation. Fourth column: Mass of a single particle. Fifth column: Total number of full-box simulations run with these common properties.}
\label{tab:fbox}
\begin{center}
\begin{tabular}{|l|c|c|c|c|}
  \hline
  Name    & $L_{box}$ & $N_p$ & $m_p$  & N\\ 
  & ($\hmpc$) &   &  ($\hmsun$) & \\
  \hline
  L5n256$^{a}$ & 5 & $256^3$ & $5.5 \times 10^5$ & 2\\
  L5n512 & 5 & $512^3$ & $6.88 \times 10^4$ & 1\\
  L25n128  & 25  & $128^3$ & $5.50 \times 10^8$ & 1 \\
  L25n256  & 25  & $256^3$ & $6.88 \times 10^7$ & 1\\
  L25n512$^{b}$  & 25  & $512^3$ & $8.59 \times 10^6$ & 3 \\
  L50n128  & 50  & $128^3$ & $4.4 \times 10^9$ & 1\\
  L50n256$^{a,c,d}$ & 50 & $256^3$ & $5.5 \times 10^8$ & 8\\
  L50n512$^{a,b,c}$  & 50  & $512^3$ & $6.88 \times 10^7$ & 3\\
  L150n256 & 150 & $256^3$ & $1.49 \times 10^{10}$&1\\
  L650n128 & 650 & $128^3$ & $9.66 \times 10^{12}$ & 1\\
  L650n256 & 650 & $256^3$ & $1.20 \times 10^{12}$ & 1\\
  L650n512$^{a}$ & 650 & $512^3$ & $1.51 \times 10^{11}$ & 2 \\
  L900n512 & 900 & $512^3$ & $3.98 \times 10^{11}$ &1\\
  \hline
  &&&& 26\\
  \hline
\end{tabular}
\end{center}
  \begin{scriptsize}
  $^{a}$ Also run at varying $\zini$\\
  $^{b}$ Tested different force resolutions\\
  $^{c}$ Adiabatic version also run using \gadget\\
  $^{d}$ Also run with \enzo
  \end{scriptsize}
\end{table}

In order to generate the initial conditions (ICs), we have used the \music{} code \citep{Hahn:2011}. The method uses an adaptive convolution of Gaussian white noise with a real-space transfer function kernel together with an adaptive multi-grid Poisson solver to generate displacements and velocities following second-order Lagrangian perturbation theory. For more specific details on the MUSIC code, we kindly refer the reader to \citep{Hahn:2011}. The transfer functions used to generate the initial conditions for this cosmology were obtained using \texttt{CAMB} \citep{Lewis:2000,Howlett:2012}. The initial redshift, $\zini$ , used for all our simulations is discussed in Section~\ref{sec:createICs}

To run the simulations in this work we used the publicly available codes \gadget, a Lagrangian TreeSPH code \citep{Springel:2005}, and \enzo, a hybrid adaptive mesh refinement (AMR), grid-based code \citep{Oshea:2004}.  We duplicate our results with these two codes not only to confirm that findings concerning the \lagvol{} hold for both approaches, but also to study the possible differences between Lagrange and Eulerian approaches to creating initial conditions for multimass simulations.  

We have run a total of 26 full-box simulations. These consist of thirteen different combinations of $L_{box}$ and particle number $N_p$ (or mass, $m_p$), summarized with an associated naming convention in Table~\ref{tab:fbox}.  The gravitational softening used for these fiducial full-box simulations was set to $\epsilon=0.02 \times L_{box}/\sqrt[3]{\npart}$. In addition, four combinations of box size and particle number (L5n256, L50n256, L50n512, and L650n512) were run with a different initial redshift $\zini$, while two cases (L25n512 and L50n512) were run with different force resolutions.  Finally, two boxes (L50n256 and L50n512) were also run with adiabatic gas physics, wherein the simulation follows a non-nonradiative hydrodynamic gaseous component, in the proportions given by the cosmological model, in addition to the dark matter.

From these full-box runs, we have selected regions of interest and run 181 zoom-in simulations, as summarized in Table \ref{tab:zooms}. The regions to zoom-in on were selected to contain a range of halo masses.  We followed \citet{Power:2003} in setting the force resolutions for the multimass simulations: $\epsilon=4\times \rvir/\sqrt{N_{\rm{v}}}$, where $N_{\rm{v}}$ is the number of particles expected in the virial radius (based on the full-box-derived virial mass and resimulated particle mass). These simulations were run with varying mass resolutions and definitions of the \lagvol{}, as discussed in the sections to follow. The mass resolution used in these runs range from an effective resolution of $512^3$ to $4096^3$ particles in the high resolution volume. We also ran sixteen adiabatic multimass simulations with \gadget{} of a Milky Way halo in which an identical softening was used for the gas particles.

\begin{table}
    \caption{Zoom-in simulations. First column: full simulation from which the resimulated halo was chosen and the \lagvol{} was calculated. Second column: Virial mass of the simulated halo at $z=0$. Third column: the number of different halos that we simulated at that mass, if more than one. Fourth column:  Number of zoom-in simulations run with \gadget{} using the cuboid \lagvol{} approach (including sixteen adiabatic simulations). Fifth column: Number of zoom-in simulations run with \enzo{} using the cuboid \lagvol{} approach. Sixth column: Total number of zoom-in simulations. These simulations include runs with varying padding, resolution, $\rtb$ defition, etc. See text for more details.}
    \label{tab:zooms}
  \begin{center}
    \begin{tabular}{|l|l|c|c|c||c|}
 	\hline
 	Full Box   & $\mvir$ ($\hmsun$) & Halos & $N_{G}$ & $N_{E}$ & Total\\ 
 	\hline
 	L5n256 & $3.55 \times 10^9$ & 1 & 3 & 0 & 3\\
 	L5n256 & $7.1 \times 10^{9}$ & 1 & 2 & 0 & 2\\
 	L5n512$^{a}$ & $3.55 \times 10^{9}$ & 1 & 10 & 0 & 10\\
 	L5n512$^{a}$ & $[5.68$-$8.52] \times 10^{9}$& 2 & 7 & 0 & 7\\
 	L5n512$^{a}$ & $2.84 \times 10^{10}$& 1 & 4 & 0 & 4\\
 	L50n512 & $[2.71$-$4.26] \times 10^{11}$& 3 & 19 & 25 & 44\\
 	L50n512$^{a}$ & $[5.68$-$8.52] \times 10^{11}$ & 23 & 88 & 2 & 90\\ 
 	L50n512  & $3.55 \times 10^{12}$ &1 & 11 & 2 & 13\\
 	L650n512 & $[5.68$-$8.52] \times 10^{14}$& 4& 8 & 0 & 8\\
       \hline
        \multicolumn{3}{|c|}{\ } &152&29&181 \\
       \hline
    \end{tabular} 
\end{center}   
    \begin{scriptsize}$^{a}$ Tested varying force resolutions\end{scriptsize}
\end{table}

\subsection{Halo Identification and Analysis}
\label{ssec:halos}

\begin{table}
\caption{Symbol definitions}
\label{tab:symbols}
\begin{center}
	\begin{tabular}{ l  l }
		\hline \hline
		Symbol & Meaning \\ 
		\hline 
		$\mvir$ & Virial mass \\
		$\rvir$ & Virial radius \\
		$\vvir$ & Virial velocity \\
		$V_{max}$ & Maximum circular velocity \\
		$c/a$ & Sphericity \\
		$\lambda$ & Spin parameter$^{a}$ \\ 
		$T$ & Triaxiality$^{a}$ \\
		$V_{max}/\vvir$ & Concentration \\
		$a_{50}$ & Time of formation \\
		$N_{subh}$ & Number of subhalos \\
		$N_{mergers}$ & Number of major mergers$^{a}$ \\
		$z_{lm}$ & Redshift of last major merger \\
		$N_{neigh}^{X}$ & Number of nearby halos$^{a}$ \\
		$D_{1,X}$ & Distance to nearest massive halo$^{a}$\\
		$M_{gas}/\mvir$ & Gas fraction$^{b}$ \\
		$\lambda_{gas}$ & Gas spin parameter$^{b}$ \\
		$\gvol$ & Cuboid \lagvol{}$^{c}$ \\
		$\chvol$ & Convex hull \lagvol{}$^{c}$ \\
		$\grvol$ & Minimum cuboid \lagvol{}$^{c}$ \\
		$\evol$ & Minimum ellipsoid \lagvol{}$^{c}$\\
		$\rtb$ & traceback radius of the \lagvol{}$^{c}$ \\
		$\ineff$ & \lagvol{} inefficiency$^{c}$ \\
		$(w/l)_{\grvol}$ & \lagvol{} cubic measure$^{c}$ \\
		$\Delta_{res}$ & Difference in resolution between the zoom \\
		 & and the parent volume ($m_p \rightarrow m_p/8^{\Delta_{res}}$) \\
		$\zini$ & Initial redshift$^{e}$ \\
		 \hline \hline
	\end{tabular}
\end{center}	
	 \begin{scriptsize}
	 $^{a}$ Further explained in Section~\ref{ssec:halos}\\
	 $^{b}$ Only applicable to adiabatic runs \\
	 $^{c}$ Further explained in Section~\ref{sec:identifying}\\
	 $^{d}$ Further explained in Section~\ref{sec:contamination}\\
	 $^{e}$ Further explained in Section~\ref{ssec:zstart}
	 \end{scriptsize}
\end{table}

We used the Amiga Halo Finder \citep[AHF,][]{Knollmann:2009} to identify halos and calculate most of the halo properties, and developed our own pipeline to calculate any property not determined by AHF. Unless specifically stated, this pipeline is based on parameters given by AHF (center, virial radius, etc). The virial mass ($\mvir$) is calculated using the overdensity ($\Delta_{\rm{vir}}$) formula from \citet{Bryan:1998} for our cosmology at each specific redshift. Our conclusions do not change when using different overdensity definitions, e.g. $\Delta_{200}=200 \rho_{crit}$. Note that we only consider distinct halos throughout this study as the \lagvol{} of a subhalo is set by its host.

\begin{figure*} 
\begin{center}
\includegraphics[width=0.45\textwidth]{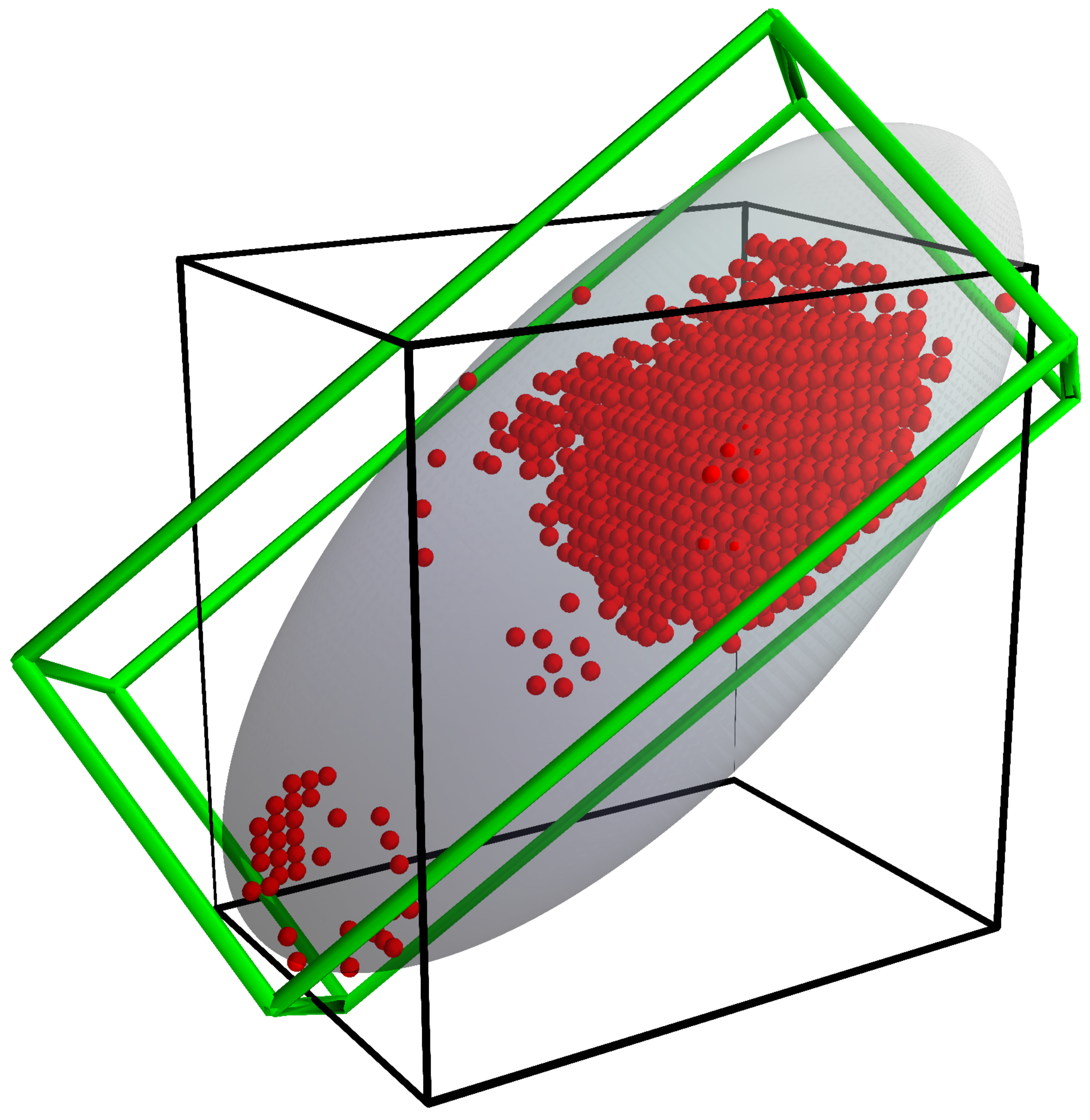}
\includegraphics[width=0.45\textwidth]{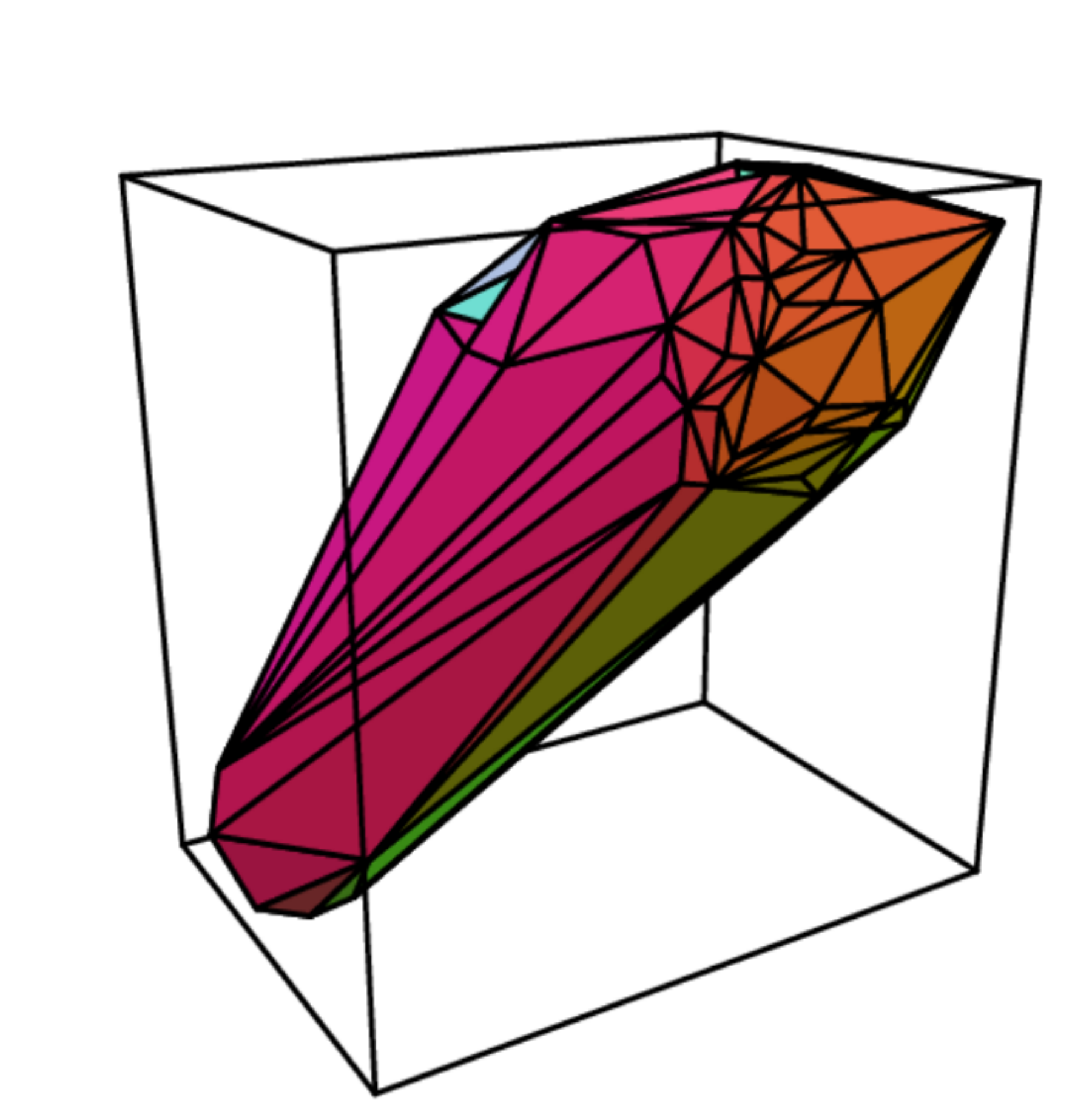}
\end{center}
\caption{\lagvol{} definitions. In the left panel, the red points indicate the $z = \zini$ positions of all the particles (1510) within $\rvir$ at $z=0$ of a halo with $\mvir=1.03 \times 10^{11} \hmsun$ (from L50n512). The black cube stands for the cuboid \lagvol{}, $\gvol$, i.e., the cuboid that is aligned with the simulation axis and that encloses all the particles. The green cuboid shows the minimum cuboid that encloses the selected particles ($\grvol$). The transparent grey ellipsoid shows the minimum ellipsoid that encloses the selected particles ($\evol$). Note the majority of the particles at $z=\zini$ are nearly planar; we find that this is a generic feature of the \lagvol{}s we explore. The right panel again shows $\gvol$ (black cube), along with the convex hull \lagvol{}, $\chvol$, for the same halo. As expected, the convex hull volume is much closer to the actual particle distribution.}
\label{fig:lagvol3D}
\end{figure*}

A list of the main halo properties used in this paper and their symbols are listed in Table~\ref{tab:symbols}. Throughout, we use the \citet{Bullock:2001} definition for $\lambda$ and the standard definition for triaxiality, $T\equiv (a^{2}-c^{2})/(a^{2}-b^{2})$ where $a$, $b$, and $c$ are the square root of the eigenvalues of the inertia tensor \citep{Allgood:2006}.  We also calculated both the number of major mergers since $z = 1$ ($N_{mergers}^{z<1}$) and over all time ($N_{mergers}^{all}$), where we define a major merger by a mass ratio greater than $0.4$ (although other values were also tested). We quantify formation time by the scale factor ($a_{50}$) or redshift ($z_{50}$) at which the virial mass of the halo reaches half of the $z=0$ virial mass.  We also calculate two environmental parameters:  $N_{neigh}^{(3)}$, following \citet{Haas:2012}, which is the number of halos outside the virial radius of a halo but within a sphere of radius $3$ Mpc from the halo center with a mass ratio greater than $0.3$ (other values for the mass ratio and the radius were also tested); and $D_{1,0.1}$, which is the distance from the center of a halo to the nearest neighbor with a mass ratio greater than $0.1$ (other values were also tested), divided by $\rvir$ of the neighboring halo.  As \citet{JeesonDaniel:2011} and \citet{Haas:2012} have shown, the first parameter correlates very strongly with virial mass, whereas the second is insensitive to mass. We find similar results as \citet{JeesonDaniel:2011} and \citet{Skibba:2011} regarding the correlations between these parameters, and therefore do not investigate these correlations further.


\section{Identifying the Lagrange Volume}
\label{sec:identifying}

Identifying the \lagvol{} that one will resimulate at high resolution is an important step in the multimass technique.  For this task, one first selects a set of particles at the redshift of interest (e.g. particles within $\rvir$, or some multiple of $\rvir$, at $z = 0$), then determines a volume that contains all of those particles at $z=\zini$ \footnote{The \lagvol{} can also be defined at $\zini=\infty$. This definition, however, makes it more difficult to relate the results to initial conditions that are actually used in practice. In Section~\ref{sec:createICs}, we further discuss the dependence of the \lagvol{} upon $z_ini$.}. However, neither step is precisely defined, and techniques used in the literature vary; here, we explore those different definitions. We first investigate the relationship between the different volumes that may be identified at $z = \zini$ for a fixed set of particles, then we discuss the effects of different methods for selecting the particle set at the redshift of interest (e.g. $z=0$).
Throughout this paper, we quote the \lagvol{} in comoving units of $h^{-3}$Mpc$^{3}$.


We consider four \lagvol{} definitions for particles:  the cuboid volume $\gvol$, the minimum rotated cuboid volume $\grvol$, the minimum ellipsoid $\evol$, and the convex hull volume $\chvol$.  The four choices are illustrated in Figure~\ref{fig:lagvol3D}.  The cuboid volume (black boxes in Figure~\ref{fig:lagvol3D}) is the minimum cuboid in ($x$,$y$,$z$) coordinates aligned with the axes of the simulation that contains the set of particles at $z=\zini$.  This is the most straightforward volume to define, set by the minimum and maximum values of $x$, $y$ and $z$ for the chosen particles. This volume is the largest of the cases we consider,
but is more practical for many codes and/or initial conditions generators.   The convex hull \citep{Barber:1996} is the minimal polyhedron that encloses the selected particles at $z=\zini$.  This choice provides the minimum volume and is illustrated relative to the cuboid in the right panel of Figure~\ref{fig:lagvol3D}.  The minimum rotated cuboid, an intermediate choice, is illustrated by the green cuboid in Figure~\ref{fig:lagvol3D}. The transparent grey ellipsoid in the same Figure that encloses all the selected particles is the minimum rotated ellipsoid \citep{Khachiyan:1979}.  Table~\ref{tab:symbols} summarizes the symbols of the main \lagvol{} parameters used in this work.

It is important to note that the convex hull is the best physical
description of the \lagvol{} and removes unphysical effects such as the
random alignment between the cuboid and the simulation axis; however,
while this alignment has no physical significance, in practice it is not
trivial to rotate the simulation and thus this issue cannot be ignored.
More complex definitions of the \lagvol{}, such as concave shapes \citep[$\alpha$-shapes][]{Edelsbrunner:1983}, have not been considered in this
work because an extra tunable parameter is required to compute them.
Though such definitions could, in principle, reduce the computational cost of a zoom-in simulation, they also require detailed studies of those parameters, and would also greatly increase the difficulty of creating the large sample of Lagrange volumes necessary for this work.

We also analyze the effect of varying the particle set that defines the \lagvol{}. One natural choice is to pick a \lagvol{} set by the particles within one virial radius of a halo.  However, such a definition may lead to contamination within the virial radius caused by wandering low resolution particles that can end up within the virial volume (see Section~\ref{ssec:effects} for a discussion of the effects of contamination); thus, one typically selects particles from a larger radius than the region of interest in the final zoom.  How conservative one must be in selecting the volume for resimulation is one of the points we return to later.    

We explore a few choices of ``traceback" radius, $\rtb$, for defining the Lagrangian region of interest. We primarily consider $\rtb$ defined as a multiple of the halo's $z=0$ virial radius; however, another option, which we explore later in the paper, is to use merger trees to track all particles that have ever been part of the halo of interest \citep{Oser:2010,Feldmann:2010}.  We apply such a method in two ways:  first, in the ``main branch" ($mb$) method, we select all particles that are within $(1, 3)\times\rvir(z)$ of the primary progenitor halo at {\em any} timestep; second, in the ``full merger tree" ($mt$) method, we select all particles that are {\em ever} within $(1, 3)\times\rvir(z)$ of {\em any halo} that ends in the $z=0$ object. For now, we will focus on the simplest definition set by the $z=0$ particle positions, but we show below that more computationally expensive approaches that rely on merger trees do not appreciably alleviate the contamination issue.


As expected, the \lagvol{} displays a strong and intuitive correlation with mass:  more massive halos have larger \lagvol{}s. To capture any unexpected trends, we therefore study the normalized \lagvol{}, $\gvol/\volvir$. 
We also investigate the inefficiency, $\ineff$, defined as the ratio between the number of particles enclosed in the \lagvol{} at $z_{ini}$ to the number of particles within $\rvir$ at $z=0$, at the same resolution.  Both parameters are well fit by identical power laws that differ only in normalization:
\begin{equation}
[\frac{\gvol^{x}}{4\pi\rvir/3},\,\ineff^x] = [A,\,B]\left(\frac{\mvir}{10^{12.0} \hmsun}\right)^{\alpha},
\label{eq:lvmass}
\end{equation}
where $x$ stands for the traceback radius definition ($\rtb=x\times\rvir$), and $A$ ($B$) gives the appropriate normalization for the normalized \lagvol{} (inefficiency). Table~\ref{tab:fitslv} presents the best fit values for $A$, $B$, and $\alpha$ for all four definitions of the \lagvol{}, for $x = 1$, $3$, and $5$.  The inefficiency parameter, although less intuitive than the normalized \lagvol{} at a theoretical level, gives a more practical value for the purpose of this work as it estimates the relative amount of resources not focused on the target halo.

\begin{figure} 
\begin{center}
\includegraphics[width=0.47\textwidth]{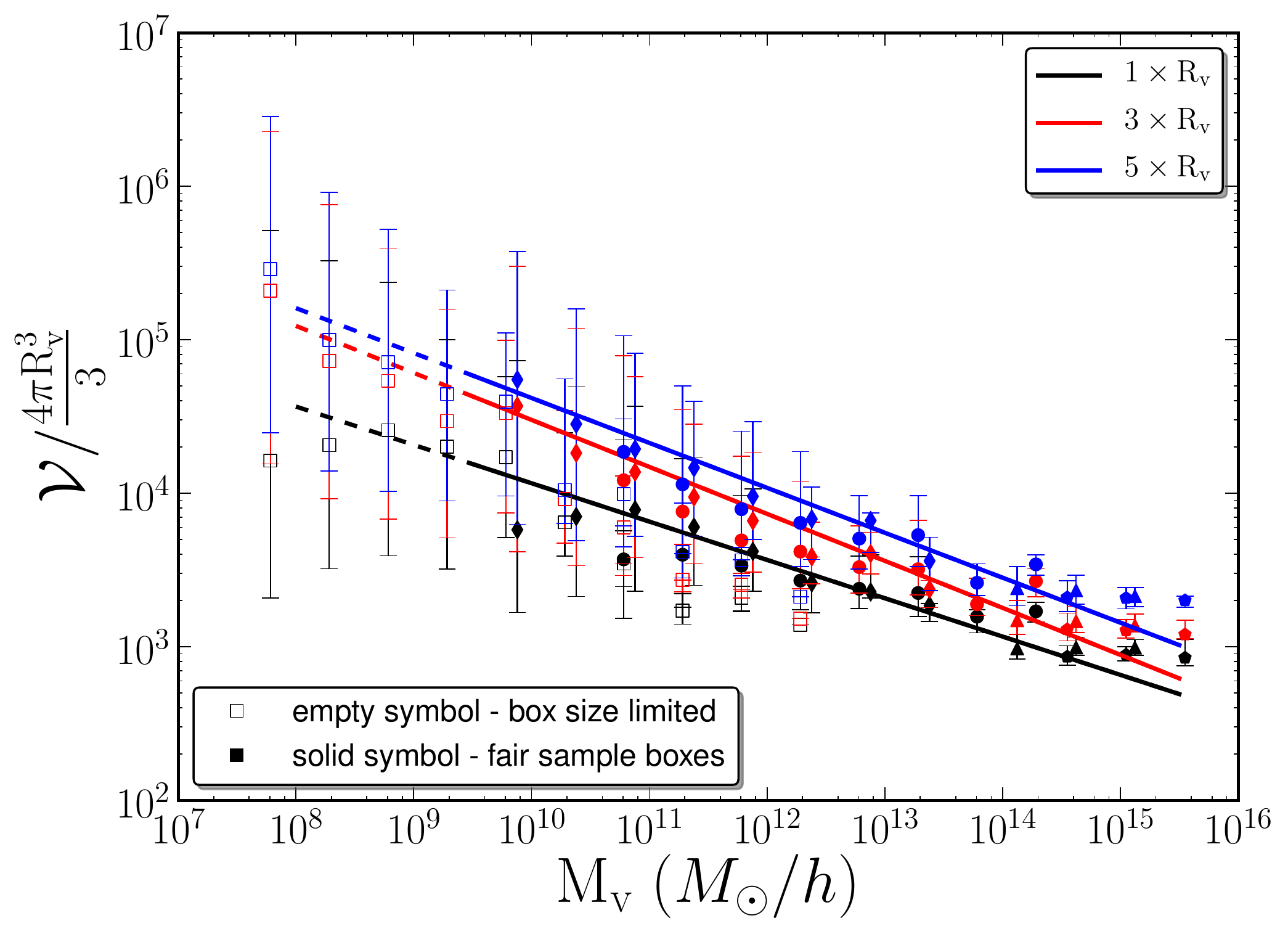}
\end{center}
\caption{Normalized \lagvol{} correlation with halo mass. We plot the normalized cuboid \lagvol{} versus virial mass for $\npart=512^3$ full-box simulations: L5n512 (empty squares), L25n512 (diamonds), L50n512 (circles), L650n512 (triangles), L900n512 (pentagons). Symbols stand for the median value of each mass bin and error bars show the 25\% and 75\% percentiles. Colors stand for the different $\rtb$ used to calculate the \lagvol{}: 1, 3 or 5 times $\rvir$. More massive halos have smaller \lagvol{}s when normalized by $\mvir$, meaning that larger halos, such as clusters, will be cheaper to run with a fixed number of particles within the halo.
The lines are the fits to the full sample, which slightly steepen with increasing $\rtb$.
Results from L5n512 are affected by the small size of the box and therefore were removed from the fits and are shown using empty squares.
The lines become dashed where the fit is extrapolated.}
\label{fig:hzvolcorrmass}
\end{figure}

Figure~\ref{fig:hzvolcorrmass} provides an illustration of how the normalized \lagvol{} scales with halo mass (and therefore also the inefficiency parameter):  we find a weak but clear anti-correlation over the entire 8 decades in $\mvir$ that our sample spans.  Thus, for a fixed number of particles within $\rvir$, multimass simulations focused on cluster halos will be computationally cheaper than galaxy halos or dwarf systems. As the slope generally steepens with increasing traceback radii, smaller halos require more resources for large zoom-volume multimass simulations. The points in Figure~\ref{fig:hzvolcorrmass} correspond to the median $\gvol/(4\pi\rvir/3)$ values at each halo mass and the error bars span the $25-75\%$ range.  Lower mass halos demonstrate more scatter in \lagvol{} at fixed virial mass.  
We will delve into correlations between the \lagvol{} and other halo properties at fixed halo mass in Section~\ref{sec:lagvol}.

\begin{table*}
\begin{minipage}{180mm}
\caption{Fits for equation~\ref{eq:lvmass} using  L25n512, L50n512, L650n512 and L900n512  simulations.}
\label{tab:fitslv}
\begin{center}
\begin{tabularx}{\textwidth}{*{13}{>{\centering\arraybackslash}X}}
    \hline 
    & \multicolumn{3}{c}{Cuboid \lagvol{}} & \multicolumn{3}{c}{Minimum cuboid \lagvol{}} & \multicolumn{3}{c}{Minimum ellipsoid \lagvol{}} & \multicolumn{3}{c}{Convex hull \lagvol{}}\\
    & \multicolumn{3}{c}{($\gvol$)}& \multicolumn{3}{c}{($\grvol$)} & \multicolumn{3}{c}{($\evol$)} & \multicolumn{3}{c}{($\chvol$)}\\
    \cmidrule(lr){2-4} \cmidrule(lr){5-7} \cmidrule(lr){8-10} \cmidrule(lr){11-13}
    $\rtb$ & $1$ & $3$ & $5$ & $1$ & $3$ & $5$ & $1$ & $3$ & $5$ & $1$ & $3$ & $5$\\ 
    \hline
 $\alpha$ &-0.25 & -0.31 & -0.29 & -0.21 &-0.26 &-0.25  &-0.22 & -0.27 & -0.26 &-0.17&-0.23&-0.22 \\
 $A$ &$10^{3.57}$ & $10^{3.87}$ & $10^{4.04}$  & $10^{3.40}$ & $10^{3.70}$ & $10^{3.87}$ & $10^{3.36}$ & $10^{3.66}$ & $10^{3.83}$ & $10^{3.04}$ & $10^{3.35}$& $10^{3.54}$\\
 $B$ &10.14 & 20.22  & 29.88 &  6.88 & 13.72 & 20.61 & 6.35& 12.45  & 18.53  &3.00 & 6.16 & 9.56\\
    \hline
  \end{tabularx}
\end{center}
\end{minipage}
\end{table*}

\begin{figure} 
\begin{center}
\includegraphics[width=0.5\textwidth]{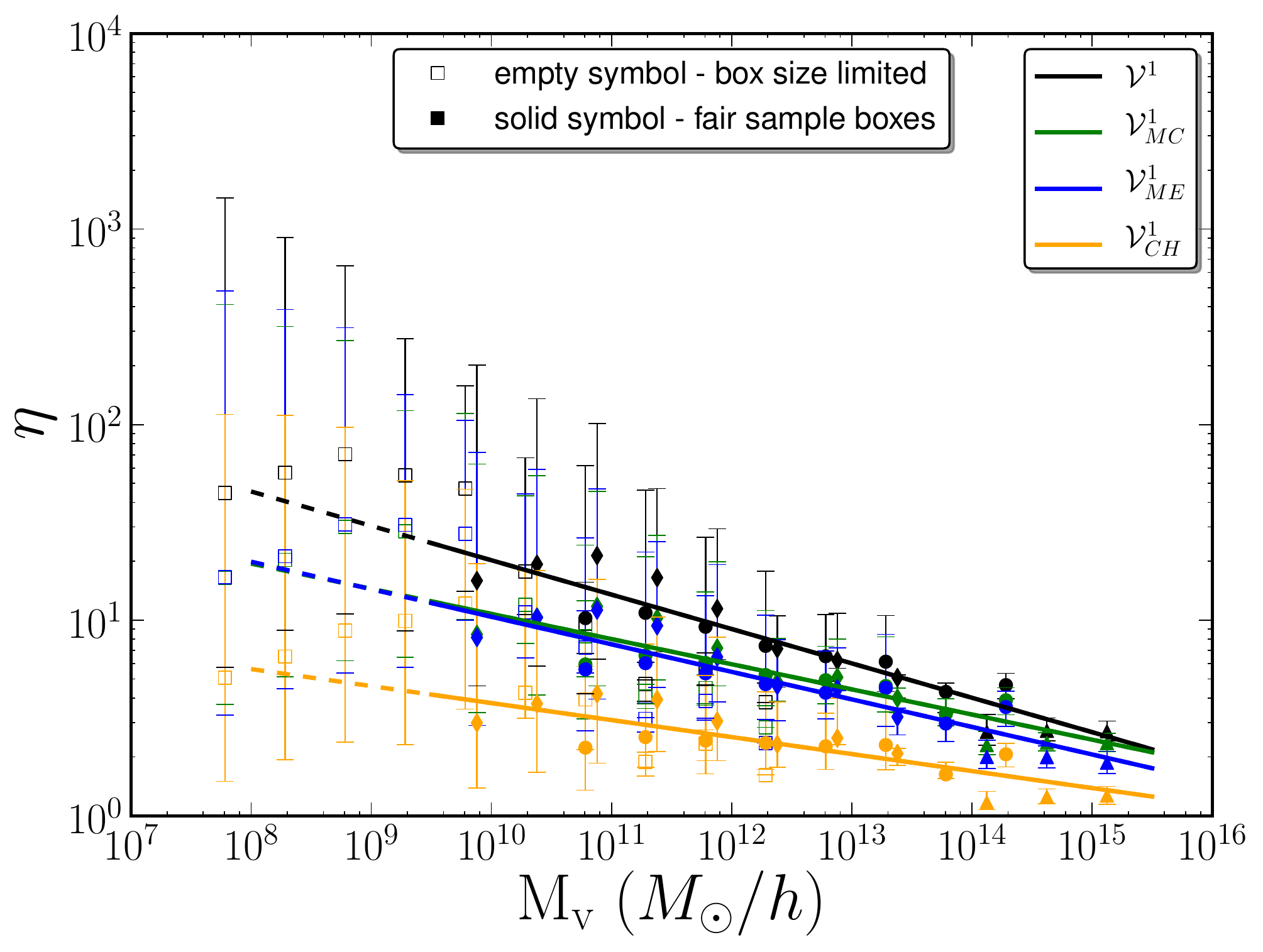}
\end{center}
\caption{Inefficiency of the \lagvol{} definitions as a function of halo mass. The \lagvol{}s are defined by the particles within a traceback radius $\rtb = 1\, \rvir$ of the halo center at $z=0$. Symbols stand for the median value of each mass bin and error bars show the 25\% and 75\% percentiles.
We plot the inefficiency, $\ineff$, of the cuboid ($\gvol^{1}$;black), rotated cuboid ($\grvol^{1}$;green), ellipsoid ($\evol^{1}$;blue) and convex hull ($\chvol^{1}$;orange) \lagvol{}s calculated in the $\npart=512^3$ full-box simulations: L5n512 (empty squares), L25n512 (diamonds), L50n512 (circles), L650n512 (triangles), L900n512 (pentagons). The lines are the fits to the filled in points; results from L5n512 are affected by the small size of the box and therefore were removed from the fits and are shown using empty squares. The lines become dashed where the fit is extrapolated.}
\label{fig:lagvolalldef}
\end{figure}

\begin{figure} 
\begin{center}
\includegraphics[width=0.5\textwidth]{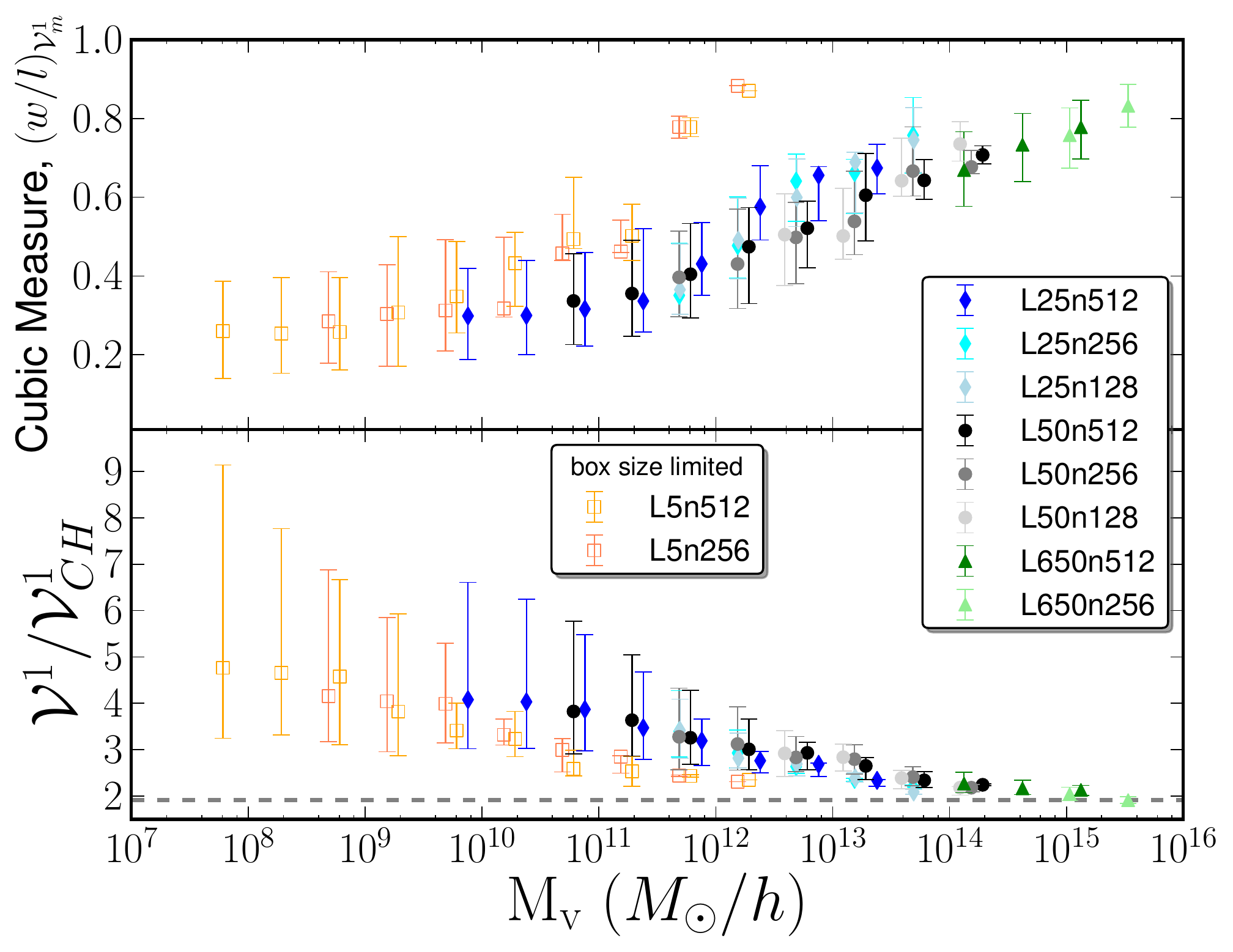}
\end{center}
\caption{Trends among halo mass and \lagvol{} shape, where particles are selected at $z=0$ within a traceback radius $\rtb = 1\, \rvir$ of the halo center (though the results are largely independent of $\rtb$).   In both panels symbols stand for the median value of each mass bin and error bars show the 25\% and 75\% percentiles. Upper Panel:  The ratio of shortest to longest axis of the minimum rotated cuboid (where a ratio of 1 implies a perfect cube) versus the virial mass.  This cubic measure of a halo's \lagvol{} is correlated with the virial mass of that halo, indicative of more massive halos having intrinsic \lagvol{} shapes that approach spheres. The higher axis ratios obtained for the most massive halos of the $L_{box}=5 \hmpc$ run (yellow and orange open points) are due to the small size of the simulation box. Lower Panel: The ratio of the simulation-box aligned cuboid \lagvol{}, $\gvol^1$, to the convex hull \lagvol{}, $\chvol^1$ (where the superscript $1$ refers to a $1 \, \rvir$ traceback radius).  The median of this ratio decreases with increasing halo mass mainly because the intrinsic \lagvol{}s are becoming spherical for the largest halos (the ratio of a cuboid of size $2l$ to a sphere with radius $l$ is marked as the dashed black line in the plot, $\sim 1.9$). The other effect that accounts for the difference between these two definitions is the misalignment of the simulation grid with the convex hull \lagvol{}. See text for more details.}
\label{fig:lagvoldef}
\end{figure}


We investigate the relationship between \lagvol{} definitions in Figure~\ref{fig:lagvolalldef}, here for $\rtb/\rvir=1$.  Plotted are the inefficiencies of the cuboid ($\gvol^{1}$), the rotated cuboid ($\grvol^{1}$), the ellipsoid volume ($\evol^{1}$), and the convex hull volume ($\chvol^1$) for a set of full-box simulations. Any purely topological \lagvol{} definition is always larger, and therefore more inefficient, than the convex hull volume, but the intrinsic scatter is large, particularly for the cuboid volume.  We also find a clear anti-correlation between both the median and the scatter in the ratio and virial mass, which can be understood by considering the shape of the \lagvol{}.  

We quantify the shape of the \lagvol{} using the ratio of the shortest to longest sides of the minimum rotated cuboid that contains the particles of interest, $(w/l)_{\grvol}$.  We plot this shape measure versus halo mass in the top panel of Figure~\ref{fig:lagvoldef}. At high halo masses, the convex hull \lagvol{} approaches a sphere, and the ratio therefore approaches that of a cube enclosing a sphere:  $6/\pi \simeq 1.9$ (shown as the gray dashed line). Lower halo masses, however, tend to have more elongated \lagvol{}s \citep{Rossi:2011}. For these halos, the angle between the simulation axes and the elongation axis can strongly inflate the value of the (unrotated) cuboid volume $\gvol$. The scatter in the bottom panel is thus dominated by the scatter in the cuboid volume, which reflects both the position of the traceback particles and their alignment with the axis of the simulation, and decreases with increasing halo mass. The bottom panel of Figure~\ref{fig:lagvoldef} confirms that the trends that we find with mass are independent of resolution.

\begin{figure} 
\begin{center}
\includegraphics[width=0.5\textwidth]{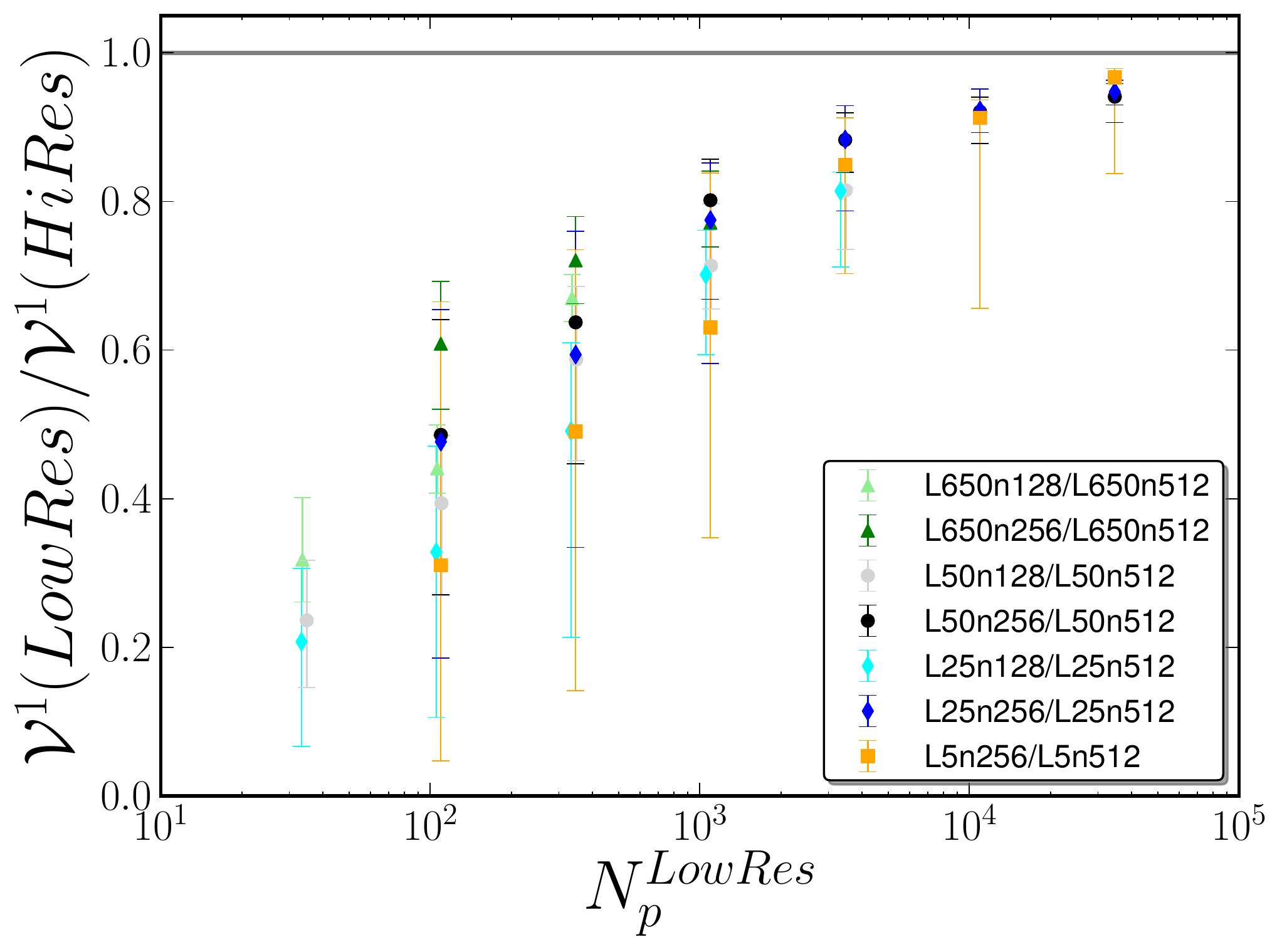}
\end{center}
\caption{Plotted is the ratio of the cuboid \lagvol{} ($\rtb = \rvir$) for identical halos, as measured at high ($\npart = 512^3$) and low  ($\npart = 128^3$ or $256^3$) resolution, versus the number of particles within $\rvir$ in the low resolution run. Symbols stand for the median value of each bin and error bars show the 25\% and 75\% percentiles.  We include halos with less than a $5\%$ variation in mass and position from four box sizes:  5 $\hmpc$, 25 $\hmpc$, 50 $\hmpc$ and 650 $\hmpc$. \lagvol{} increases with increasing resolution --- i.e. \lagvol{} does not converge, even for halos with tens of thousands of particles within $\rvir$ in the low resolution run. For cases where $\npart \sim 100$, the \lagvol{} can easily be underestimated by $\sim50\%$.}
\label{fig:lagvolres}
\end{figure}

Due to the nature of the \lagvol{}, a single outlying particle can strongly alter the cuboid \lagvol{}; as such, it is  important to ensure that the \lagvol{} is converged. Figure~\ref{fig:lagvolres} shows the ratio between the \lagvol{} of {\it identical} halos as measured at two resolutions -- only $z=0$ halos where the position of their center varies by less than 5\% of their virial radius and whose masses also vary by less that 5\% between the two resolution runs are plotted.\footnote{When comparing simulations with different resolutions, once must rely on the halo catalogs, rather than direct particle correlations, to identify matching objects. We search for halos with similar positions and masses between the catalogs; however, this process is not straightforward as both simulations may have slightly different populations of halos due to resolution effects. The 5\% cut avoids incorrect matches which, although they do not alter the general trends, introduce unnecessary noise.}.  We see that, for both $\gvol$ and $\chvol$, the low resolution simulation {\em always} underestimates the \lagvol{}, even for halos with more than $10^4$ particles. By extrapolating the results of Figure~\ref{fig:lagvolres}, one could be tempted to obtain the value at which the \lagvol{}s show convergence (i.e. a ratio of $1.0$). However, such an extrapolation relies on the assumption that the Lagrange volume computed at the $512^3$ effective resolution has no errors, which we will show is not true.  Regardless, it is clear that one must correct for this effect when building zoom-in initial conditions at higher resolutions.  This correction may be small if the Lagrange volume is computed with a large number of particles, but becomes increasingly important as the number of particles used decreases.  There is also a second-order mass dependence, which is directly related to the anti-correlation presented in Figure~\ref{fig:hzvolcorrmass} and also observed in Figure~\ref{fig:lagvolres}:  at the same resolution, dwarf galaxy halos are less converged than cluster halos.

This result may initially be quite surprising.  It seems unlikely that, for halos with more than $10^4$ particles in the low resolution run, the \lagvol{} is so poorly defined while all other halo properties are in excellent agreement with higher resolution simulations. Again, this can be understood by considering the increasingly irregular shape of the \lagvol{} at increasingly higher mass resolutions.  Regardless of the method for determining the \lagvol{}, it ultimately depends only on a few particles --- in the case of the cuboid \lagvol{}, for example, a maximum of six. As the number of particles (resolution) increases, we better sample the \lagvol{}, selecting particles from farther away that migrate into the halo. The contribution of these particles to the halo mass is insignificant, so this and other halo properties are not strongly affected, but, as we show here, it prevents full convergence of the \lagvol{} and, as we will show in Section~\ref{sec:contamination}, can be important for hydrodynamic runs. We have also checked that this effect is not due to two-body scattering effects. As the particle mass decreases and the force resolution increases, it may become increasingly easier for particles to scatter drastically and end up somewhere close to the halo. We therefore compare the \lagvol{}s obtained from one of our standard runs (L50n512) to an otherwise identical run that uses the softening lengths used in the next lower resolution level (i.e., L50n256). We find that there are not significant variations or trends in the ratio of the \lagvol{}s between these runs.

\begin{figure} 
\begin{center}
\includegraphics[width=0.5\textwidth]{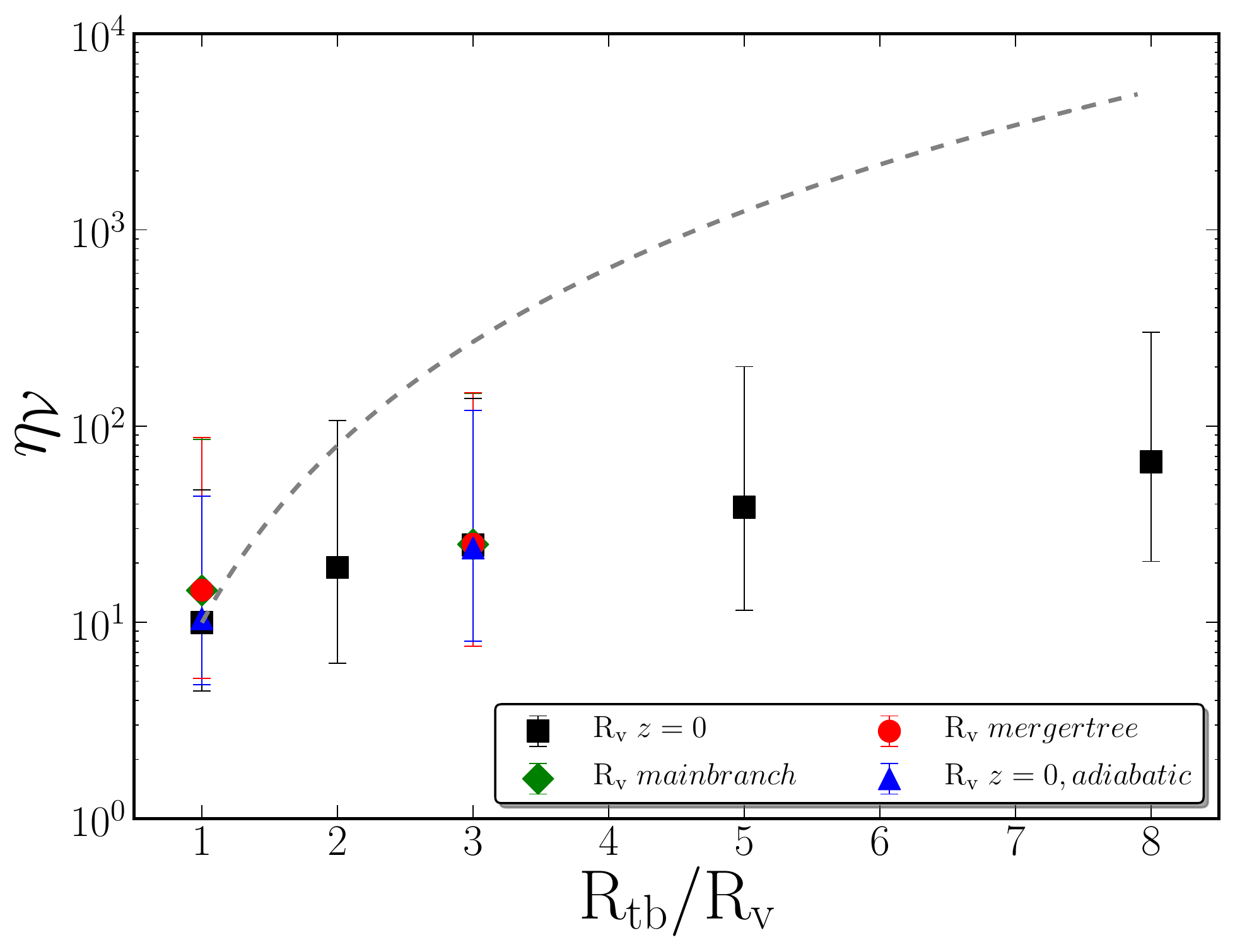}
\end{center}
\caption{Plotted is the inefficiency of the cuboid \lagvol{}, $\ineff_{\gvol}$, as a function of traceback radius $\rtb$. The median values are shown for all halos in the L50n512 simulation with $\npart\geq500$ (excluding subhalos) and error bars stand for the $25\%$ to $75\%$ percentile. The main branch \lagvol{} $\rtb$ definition, i.e., following the evolution of the halo and selecting all particles that are ever within X $\times \rvir(z)$ of the main halo. $\rvir^{mt}$ stands for the full merger tree \lagvol{} $\rtb$ definition, i.e., considering the full merger tree. Finally, $\rvir^{ad}$ stands for the \lagvol{} computed for the adiabatic version of the same initial conditions. The dashed grey line indicates how the inefficiency of the \lagvol{} would increase if $\lvol \propto \rtb^{3}$.}
\label{fig:xrvir_vrs_Lvol}
\end{figure}

These results motivate selecting a \lagvol{} for a multimass simulation using $\rtb > \rvir$.  Figure~\ref{fig:xrvir_vrs_Lvol} shows the cuboid \lagvol{} for traceback radii between $\rvir$ and $8 \, \rvir$, normalized by the virial volume, as a function of $\rtb$.  We note that the \lagvol{} grows rather slowly with $\rtb$ --- a traceback radius of $\rtb=5 \, \rvir$ increases the $z=0$ traceback volume by $125$, but the average \lagvol{} in the initial conditions increases by only $\sim2$.  Thus, the extra computational cost associated with a larger traceback radius is significantly less than one may have expected. 

Figure~\ref{fig:xrvir_vrs_Lvol} also includes results for the merger tree techniques and the adiabatic runs.  In the case of the former, we find that they are generally equivalent to choosing a larger $\rtb$ at $z = 0$. In particular, all $\gvol^{1,mb}$ and $\gvol^{1,mt}$ \lagvol{}s are enclosed by the $\gvol^{2}$ choices. In the case of $\gvol^{3,mb}$ and $\gvol^{3,mt}$, there is no significant difference with $\gvol^{3}$. As the merger tree analyses take a significant amount of computational effort with no obvious benefit, we will consider only the case of selecting traceback particles via a radius defined at a single snapshot from here on. Results from the adiabatic runs are not significantly different from those computed for the same halos in a dark matter-only simulation.
In the following section, we explore how contamination affects a zoom-in halo with using different $\rtb$.

\section{Contamination}
\label{sec:contamination}

One concern with any zoom simulation is that the region of interest be free from contamination by low resolution particles throughout cosmic time.  Naively, one might think this can be easily accomplished by simply choosing a \lagvol{} slightly larger than the high resolution region; however, no established approach exists in the literature --- instead, each group uses its own recipe with varying degrees of success.  In this section we investigate this issue systematically using several multimass simulations created with cuboid \lagvol{}s.  We will later discuss the differences between this definition and that using the convex hull volume, but it is important to state here that the main conclusions of this work are valid for both definitions.  We primarily focus on Milky Way-mass halos, but we also explore a few clusters and dwarf galaxy halos (see Table~\ref{tab:zooms}).

\begin{figure} 
\begin{center}
\includegraphics[width=0.47\textwidth]{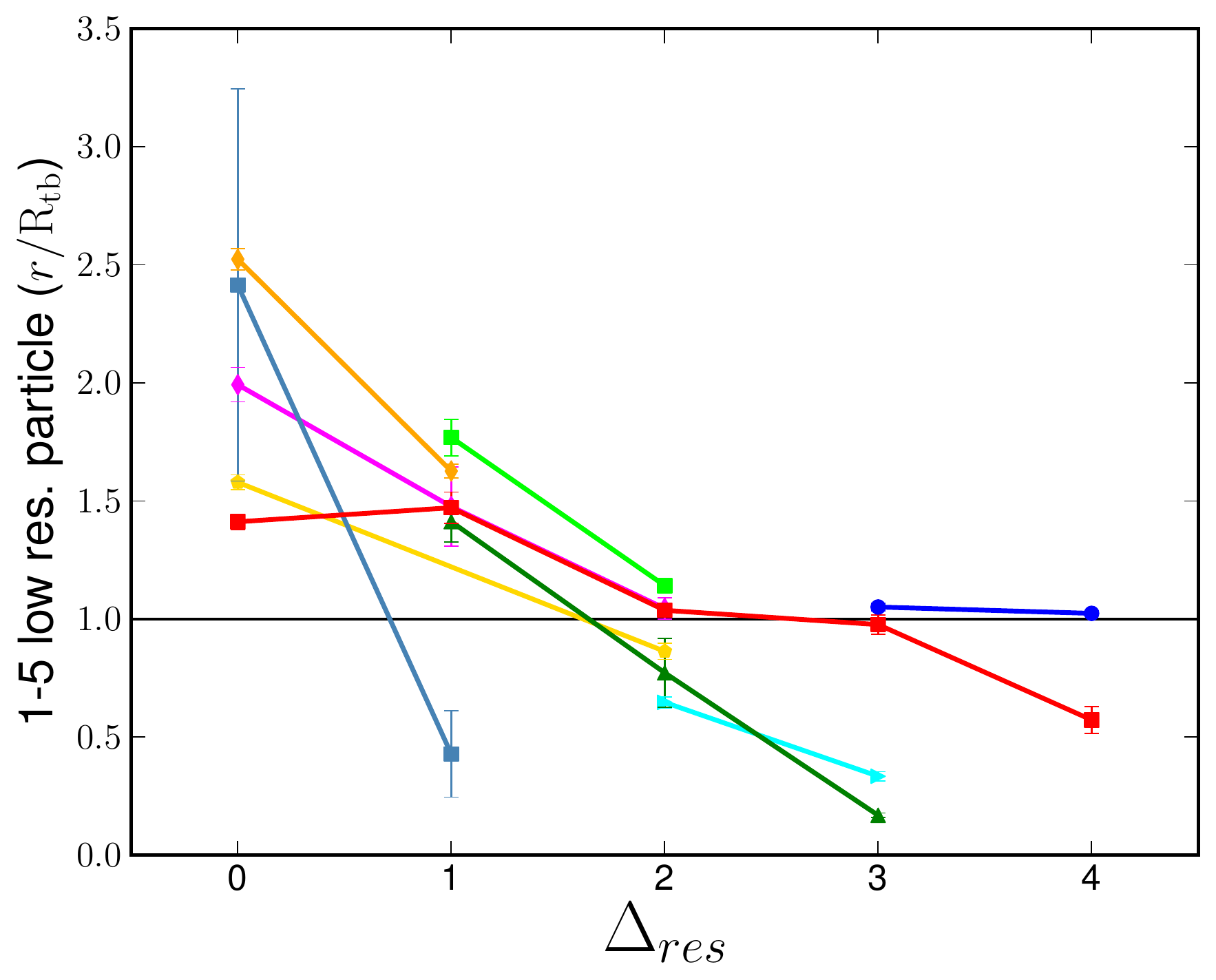}
\end{center}
\caption{Contamination in multimass simulations. Plotted is the ratio between the distance where low resolution particles are found and the traceback radius used to select the particles ($\rtb$) as a function of resolution steps (i.e. factors of $2^3$ in $\npart$) between the high resolution region and the simulation used to calculate the \lagvol{}. Each symbol represents a different zoom-in simulation. The error bars indicate the range in distance between the first and fifth low resolution particle closest to the center of the region of interest and the symbol is located at the mean value of these two. Lines connect symbols which represent zoom-in simulations with identical \lagvol{}s but varying the mass resolutions. 
Naively, one might expect the y-axis ratio to hover around $1$, which would indicate that particles in multimass simulations end in roughly the same location as in full-box runs. However, we find that for a fixed \lagvol{}, low resolution particles wander closer to the center of the region of interest as the number of resolution steps increases.}
\label{fig:contamination1}
\end{figure}

Figure~\ref{fig:contamination1} illustrates one of the main conclusions of this work: the chance for contamination increases with the level of zoom.  The vertical axis shows the distance from the host halo center to the first low-resolution particles for several of our zoom simulations.  Particle distances are plotted in units of the track-back radius $\rtb$ and shown as a function of resolution steps $\Delta_{res}$ between the full box used to calculate the \lagvol{} and the zoom simulation itself.  Each $\Delta_{res}$ step corresponds to a factor of $2^3$ increase in effective particle number,  such that the mass resolution improves as $m_p \rightarrow m_p/8^{\Delta_{res}}$.     In ideal cases, there would be no points with $r/\rtb < 1$.   Error bars indicate the range in distances between the first and fifth low resolution particles closest to the center of the halo.   Lines connect zoom-in simulations with identical \lagvol{}s but with varying the mass resolution.   We see that as the resolution difference between the full-box and multimass simulations increases, the likelihood that low resolution particles wander into the region of interest grows quickly.  If the same \lagvol{} is used to simulate the same halo at increasing resolution, contamination becomes more likely.   It is clear that if one is concerned about contamination within $\rvir$ then setting $\rtb = \rvir$ is problematic even for $\Delta_{res} = 1$ (i.e. a factor of $2^3$ in mass).   More importantly, to ensure that the region within $\rvir$ is uncontaminated,  $\rtb$ must increase systematically with $\Delta_{res}$.

\subsection{Avoiding contamination}
\label{ssec:formula}

Given that the \lagvol{} of a halo can be irregularly shaped, its overall size can be affected by a few particles.  As we have seen, the \lagvol{} of an individual halo will often grow when resampled at higher resolution.  However, even if there is no change in the particle mass, small numerical differences from run to run can result in slightly different trajectories through time.   Of course, it is possible to choose a small \lagvol{} and, purely by luck, have no contamination in the region of interest. However, this is not wise given the computational cost of many zoom simulations.

\begin{figure} 
\begin{center}
\includegraphics[width=0.48\textwidth]{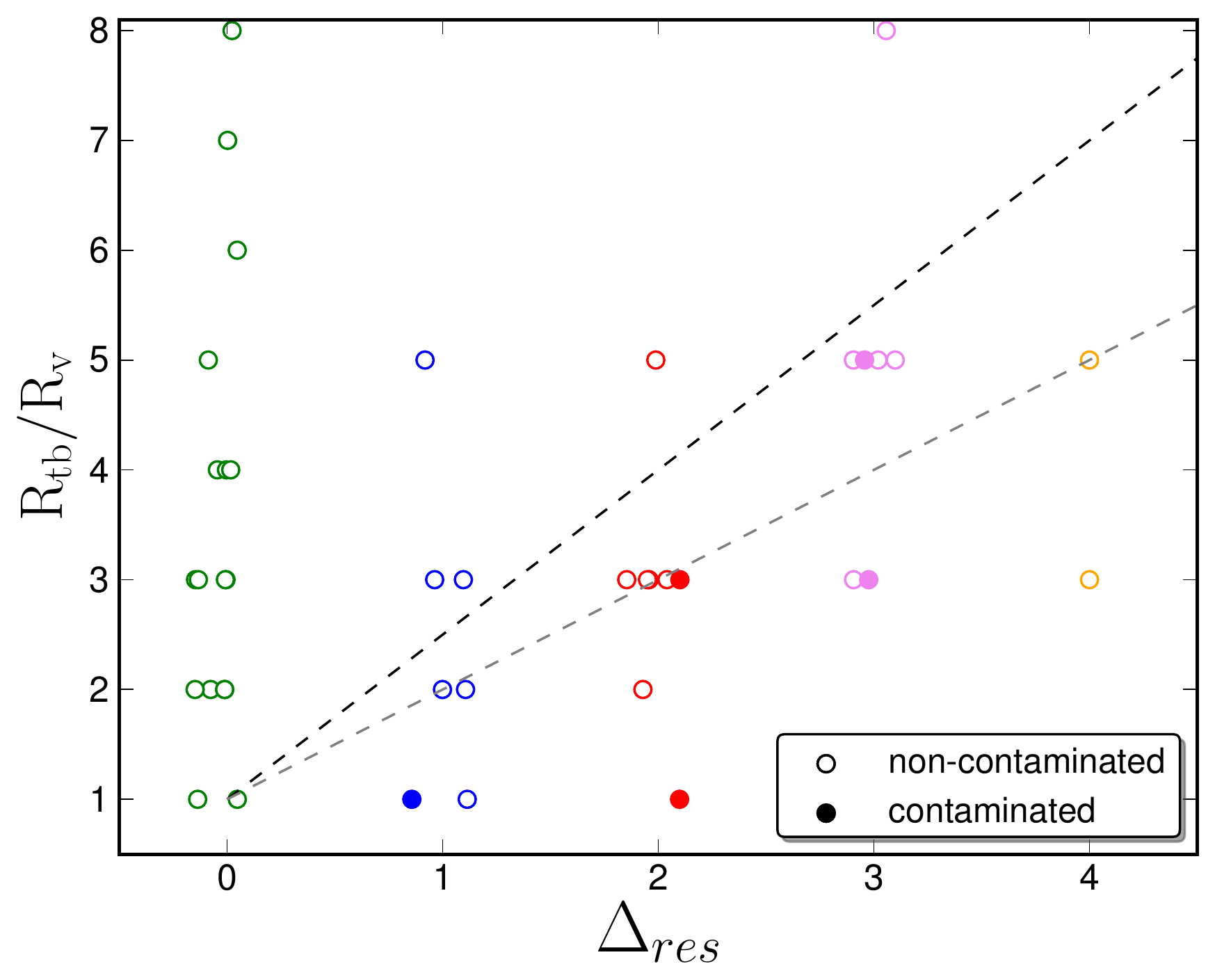}
\end{center}
\caption{Results from multimass simulations used to give recommendations on the size of the cuboid \lagvol{}. For each zoom-in simulation the size of the \lagvol{} used, $\rtb$, versus the $\Delta_{res}$ is plotted. Open symbols stand for non-contaminated halos and filled symbols for contaminated ones. The black dashed line stands for equation~\ref{eq:lvol} (slope of $1.5$), while the grey dashed line indicates an identical relation but with a slope of $1$.}
\label{fig:zoomsEQ}
\end{figure}

We present an empirical relation for defining a \lagvol{} that guarantees a resimulated halo is free from contamination within $\rvir$.  
Specifically,  for a simulation that zooms by a factor of $\Delta_{res}$ ($m_p \rightarrow m_p/8^{\Delta_{res}}$) compared to the initial box, then the \lagvol{} can be defined using the particles within a sphere around $z=0$ particles of radius 
\begin{equation}
\rtb = (1.5 \, \Delta_{res}+1)\times \rvir \, ,
\label{eq:lvol}
\end{equation}
in order to conservatively avoid halo contamination by low res particles. As discussed above, if we increase the resolution of the zoom-in simulation we need to increase the \lagvol{}. This formula is based on the results of the $\sim130$ multimass simulations we have studied for this work, all of which used \lagvol{}s calculated from at least $4000$ particles. This formula is expected to hold just between the tested range of $\Delta_{res}$ studied in this work, $0$ to $4$. Figure~\ref{fig:zoomsEQ} shows a set of these simulations in which empty circles indicate non-contaminated halos and filled circles signify contaminated runs. The black dashed line stands for equation~\ref{eq:lvol} (slope of $1.5$) and grey dashed line stands for the same equation but with a slope of $1$ to guide the eye. 

We stress here that Equation~\ref{eq:lvol} is derived from experimental results using full-box simulations with resolutions up to $512^3$ particles and zoom-in simulations with effective resolutions as high as $4096^3$.  The relation is primarily driven by the correlation between the \lagvol{} and the resolution at which it is calculated.  As pointed out above, we expect that the \lagvol{} will converge when it is calculated using a very high number particles; thus, if the \lagvol{} is initially constructed from a relatively high resolution simulation, Equation~\ref{eq:lvol} may be significantly relaxed.  Conversely, if the \lagvol{} is calculated with fewer than $\sim4000$ particles, it should be greatly increased.  Depending on the desired $\Delta_{res}$, it may even be computationally cheaper to use intermediate resolution runs to refine the \lagvol{} for the highest resolution (and therefore most expensive) simulation.  

There is also a second-order mass dependence that we do not consider, which is directly related to the anti-correlation presented in Figure~\ref{fig:hzvolcorrmass}.  However, to fully capture this effect, a much larger sample of multimass simulations is needed.  Here we simply note that zoom-in simulations aimed at dwarf-sized halos require proportionally larger \lagvol{}s than those aimed at clusters.

\subsection{Effects of Contamination}
\label{ssec:effects}

 \begin{figure*} 
\begin{center}
\includegraphics[width=0.33\textwidth]{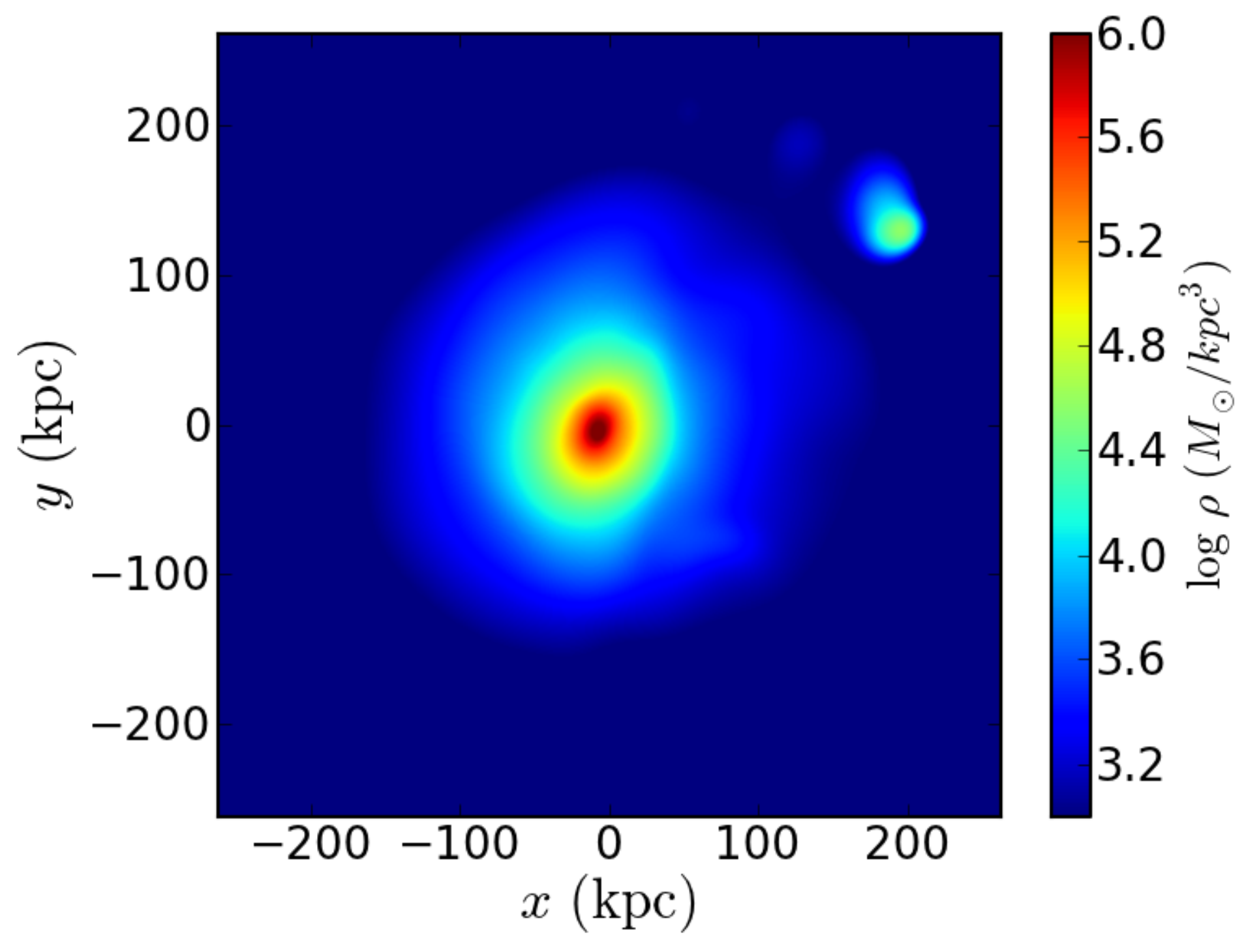}
\includegraphics[width=0.33\textwidth]{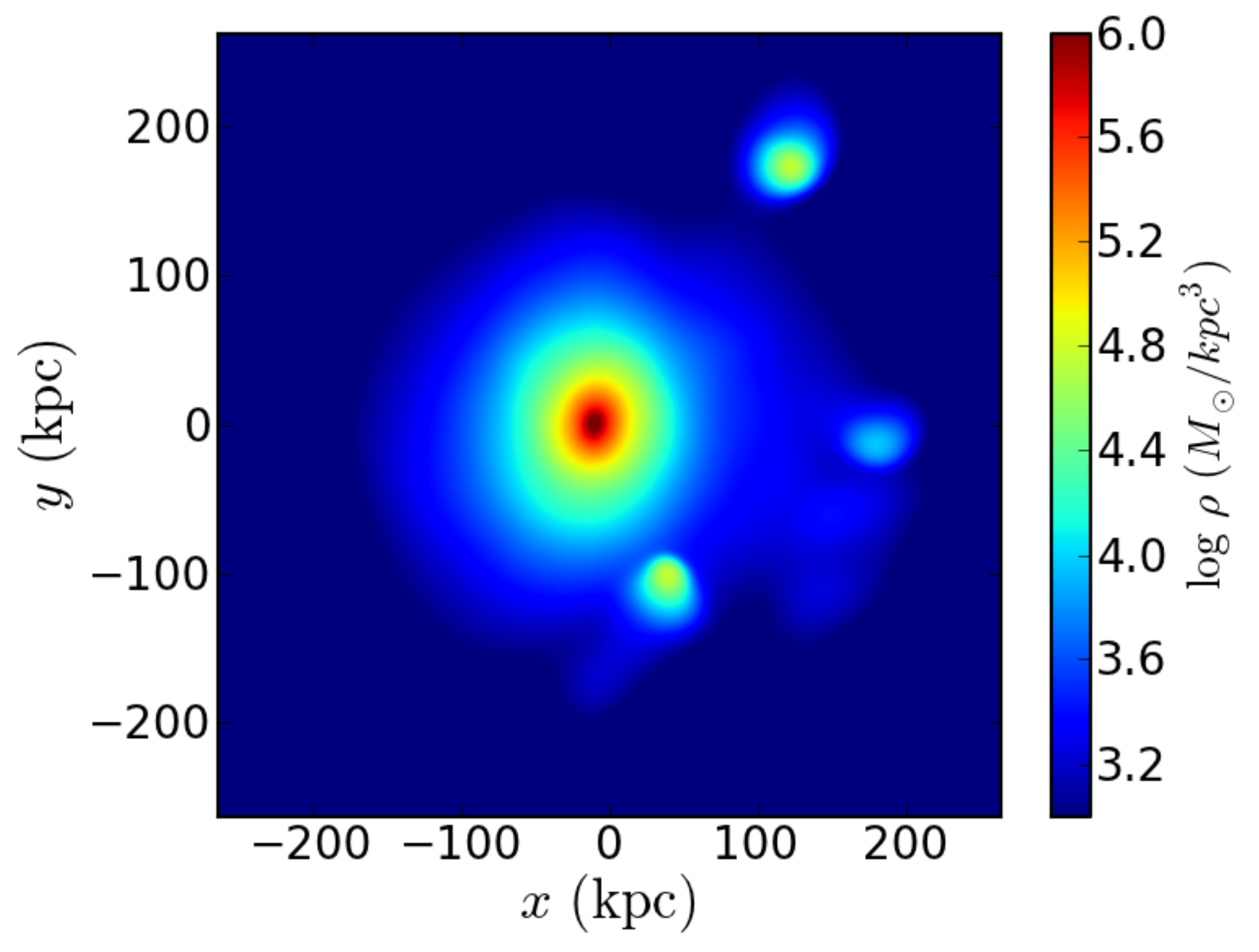}
\includegraphics[width=0.33\textwidth]{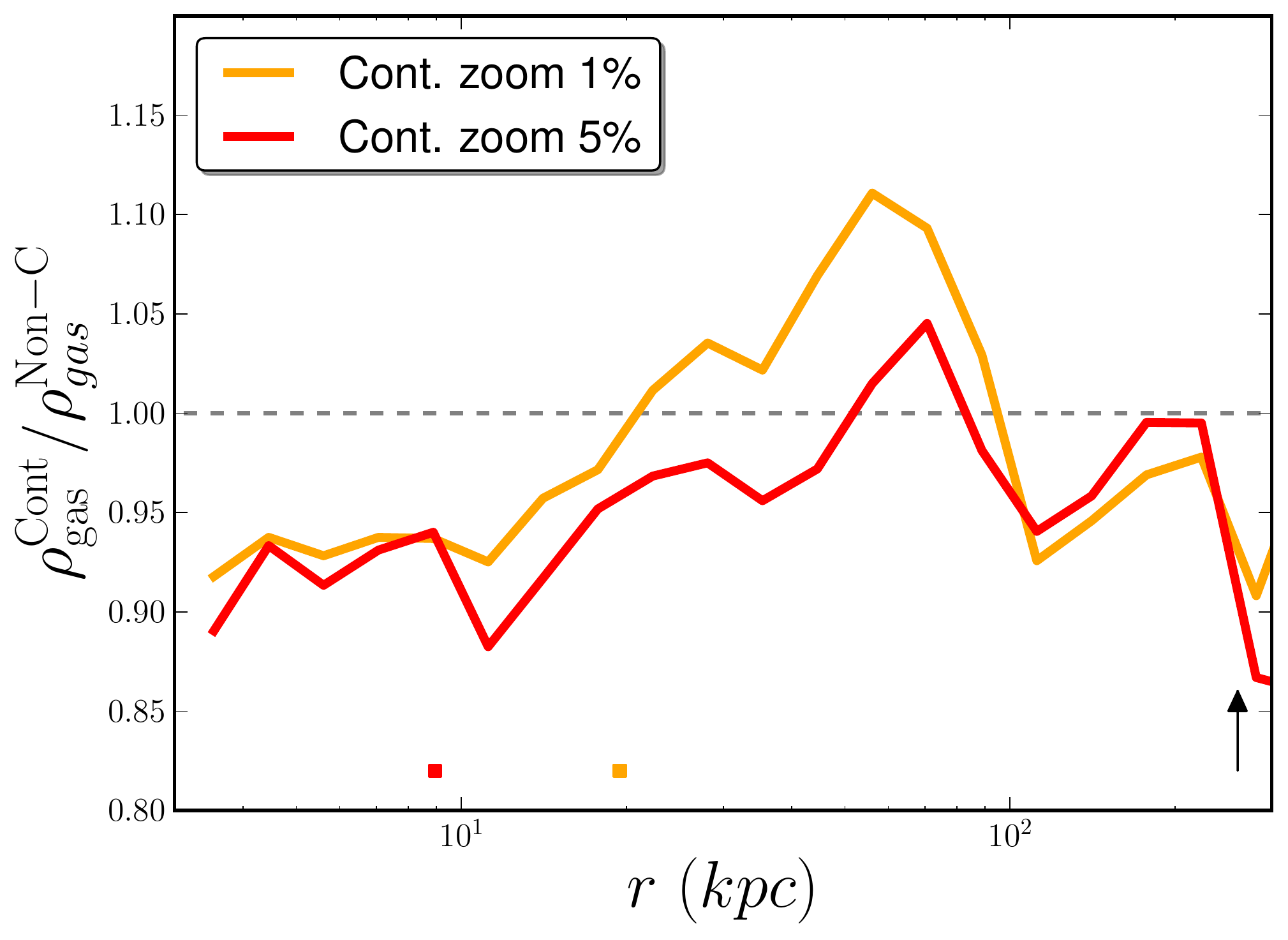}
\end{center}
 \caption{Contamination effects.  
 Left and middle panels: The left panel shows the gas density of a non contaminated \gadget{} adiabatic run (effective resolution of $1024^3$ in a $L_{box}=50 \hmpc$ box run) of a MW halo ($\rvir\sim260$ kpc). The middle panel shows a 5\% mass contaminated run, showing how dark matter low resolution particles can act as an artificial sink for gas particles. 
 This effect is less significant for a 1\% mass contaminated run, but likely it will be more relevant once cooling and star formation have been introduced.  Right Panel: Comparison of the gas density profiles between contaminated and non-contaminated zoom-ins of the same MW halo. The ratio relative to the non-contaminated run is shown for the 1\% contaminated halo (orange) and the 5\% contaminated halo (red). The squares mark the smallest radii at which low resolution particles are found at $z=0$.  Contaminated runs under-predict the gas density within the radius of contamination --- a result that also holds in multimass simulations with effective resolutions of $512^3$. See text for more details.}
\label{fig:testzoompropGbarprof}
\end{figure*}

The previous section discusses in how contamination can be prevented; however, one might argue that contamination on the order of a few percent does not matter.  Clearly, large-scale properties (e.g. halo mass) are mostly unaffected by a small number of low resolution particles within the virial radius.  However, many zoom-in simulations are performed with the goal of determining  small-scale features within the halo or for gas dynamical simulations that are sensitive to the existence of contaminating massive particles.  Here we explore whether even a small number of contaminating low resolution particles can alter halo properties.

We have found that for dark matter-only simulations, no halo properties (shape, concentration, spin, etc.) show significant differences, provided the contamination remains less than $2\%$ of the mass within the virial radius.  For adiabatic multimass simulations, however, we find that contamination has a significant effect on the gas properties of the host halo.  As an example, we plot the gas and dark matter density profiles of a specific halo (drawn from the L50n512-adiabatic box) in Figure~\ref{fig:testzoompropGbarprof}; lower panels show the deviation between contaminated and uncontaminated zoom-in runs, where $\Delta_{res}=2$.  The three multimass simulations are identical aside from their \lagvol{}s.  We find that when low resolution particles are present within $\rvir$, the gas density profile is significantly altered from the uncontaminated case. Such variation is evident in all contaminated adiabatic zoom-in simulations that we have run, and we suggest therefore that contaminated halos will exhibit systematically lower gas densities. Additionally, the contamination within a halo correlates with the total gas mass within $\rvir$, and therefore the baryon fraction:  $5\%$ contamination results in a baryon fraction $5\%$ smaller than that measured in an uncontaminated halo.  We also note that low resolution dark matter particles within $\rvir$ tend to act as sinks for gas particles, leading to artificial substructure formation.  This effect is small but appreciable in our adiabatic test runs; including cooling and star formation should act to exacerbate such numerical artifacts.

Finally, we investigate possible edge effects in our adiabatic simulations.  Typically only the high resolution region in a baryonic Lagrangian zoom-in simulation is populated with gas particles; therefore, an artificial boundary condition is created at the edge of the high resolution region wherein the gas density instantaneously drops to zero.  Eulerian zoom-in simulations use a low resolution grid outside of the high resolution region and therefore have no sharp edge.  Though one may worry that such an edge creates unphysical effects in the gas properties, we find no relevant edge effects as the profiles only diverge at well beyond low resolution particles.  Any edge effects, therefore, are completely overwhelmed by errors introduced by contamination.

\section{Selecting a Halo for Resimulation}
\label{sec:lagvol}

As Figures~\ref{fig:hzvolcorrmass} through \ref{fig:xrvir_vrs_Lvol} show, the scatter in the \lagvol{} of a halo at fixed mass is large, regardless of the volume definition. Larger \lagvol{}s are necessarily more expensive computationally; choosing a halo with a small \lagvol{} is the easiest way to reduce the required memory and CPU time.  However, choosing halos with small \lagvol{}s may bias one towards specific halo properties.  In this section, we explore the correlations between \lagvol{} and other halo properties.  We focus on the \lagvol{} as defined using $\rtb=3\rvir$, but all trends here can be extrapolated to different traceback radii.

\subsection{Lagrange Volume Correlations}
\label{ssec:lv_corr}

\begin{figure*}  
\begin{center}
  \includegraphics[width=0.7\textwidth]{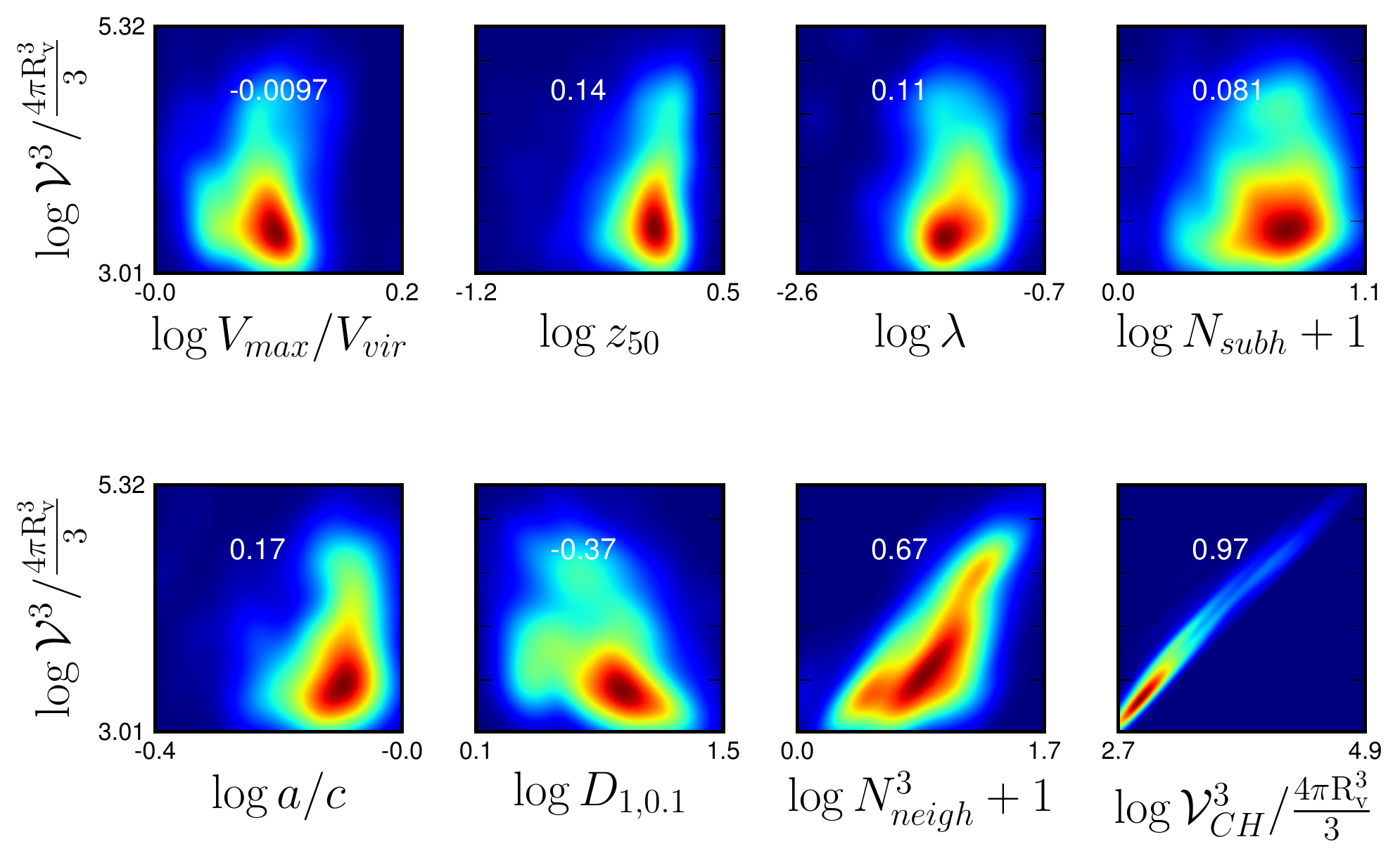}
\end{center}
 \caption{The relation between \lagvol{} and several halo properties for all halos with $10^{11.8} < \mvir/M_{\sun} < 10^{12.2}$. The halo properties from left to right are concentration, formation time, spin, substructure, shape, environment, and the convex hull \lagvol{}. The numbers in each panel give the Spearman correlation between that property and $\gvol$.  Only the environmental parameters exhibit relevant correlations with both types of the \lagvol{}. Color maps stand for the probability density function obtained for each two variables, where dark red marks the peak of the distributions and dark blue the lowest value.}
\label{fig:hzvolcorrbin}
\end{figure*}

We study the relationships between the different definitions of \lagvol{} and other halo properties mentioned in Section~\ref{sec:identifying}.  We have created a sample of well-resolved host halos that spans 8 decades in halo mass by combining all full-box simulations with $\npart = 512^3$ (L5n512, L25n512, L50n512, L650n512 and L900n512).  As discussed in Section~\ref{sec:identifying}, $\gvol$ correlates strongly with $\mvir$; however, the normalized \lagvol{} is anti-correlated with the virial mass --- more massive halos have smaller normalized \lagvol{}s (Figure~\ref{fig:hzvolcorrmass}, Spearman correlation coefficient of $\sim0.6$).  Consequently, the normalized \lagvol{} exhibits the same correlations with other halo parameters as the virial mass \citep{JeesonDaniel:2011}.

However, when initializing a multimass simulation, one typically has a specific halo mass in mind.  We thus investigate whether the \lagvol{} correlates with any other halo properties (See Table 3) at fixed virial mass.  Remarkably, aside from the expected (near-perfect) correlation with $\chvol$,  the only correlations we find with $\gvol$ (with Spearman coefficient greater than 0.3) involve environmental measures.  This is illustrated in Figure~\ref{fig:hzvolcorrbin}, which shows the correlation between $\gvol$ and other halo properties for a narrow mass range of halos ($10^{11.8} < \mvir/M_{\sun} < 10^{12.2}$), with the Spearman correlation listed in each panel. In particular we show relationships between $\gvol$ and concentration ($V_{max}/\vvir$), formation time, spin, number of subhalos, sphericity, environment (as quantified by $D_{1,0.1}$ and $N_{neigh}^{3}$ --- see Section~\ref{ssec:halos}), and the convex hull \lagvol{}.  

The correlation between \lagvol{} and environment is notable.  One would expect that the highest density regions would show larger $\gvol$ values, given that these regions are likely more chaotic at high redshift, resulting in a greater likelihood that initially outlying particles will accrete onto the halo.  Interestingly, however, at almost all environments (except for the densest of all) there remains quite a spread in halo \lagvol{}s.  One implication is that choosing an isolated halo for resimulation does not guarantee a small \lagvol{}.

In summary, we find that at fixed virial mass, a halo's formation history, structure, and dynamic state is largely independent of its  \lagvol{}.  We would expect to see correlations with the halo's formation history and thus other halo properties if we instead examined the density fluctuations of matter in the \lagvol{} \citep{Lee:2009,Ludlow:2011}, but as we have noted before, the size of the \lagvol{} is defined by the small subset of particles that travel the farthest to join the halo. These particles contribute very little to the total mass and thus, for a fixed mass, the \lagvol{} of a halo is unrelated to the exact nature of that halo's collapse.
We do note, however, that the \lagvol{}, and therefore the number of high resolution particles included in the ICs, is not the only parameter that sets the final cost of the simulation.  In particular, a halo that undergoes a low redshift merger is likely to have a higher cost:  higher density regions require a lower integration timestep, and late mergers produce an increase of density for a significant number of particles.  However, these and other sources of CPU cost are independent of the \lagvol{} and, moreover, smaller \lagvol{}s {\em necessarily} require less memory and disk space; therefore, we recommend minimizing the \lagvol{} in all cases.

\subsection{Choosing Halos to Resimulate}
\label{ssec:b.2}

\begin{figure*} 
\begin{center}
\includegraphics[width=0.32\textwidth]{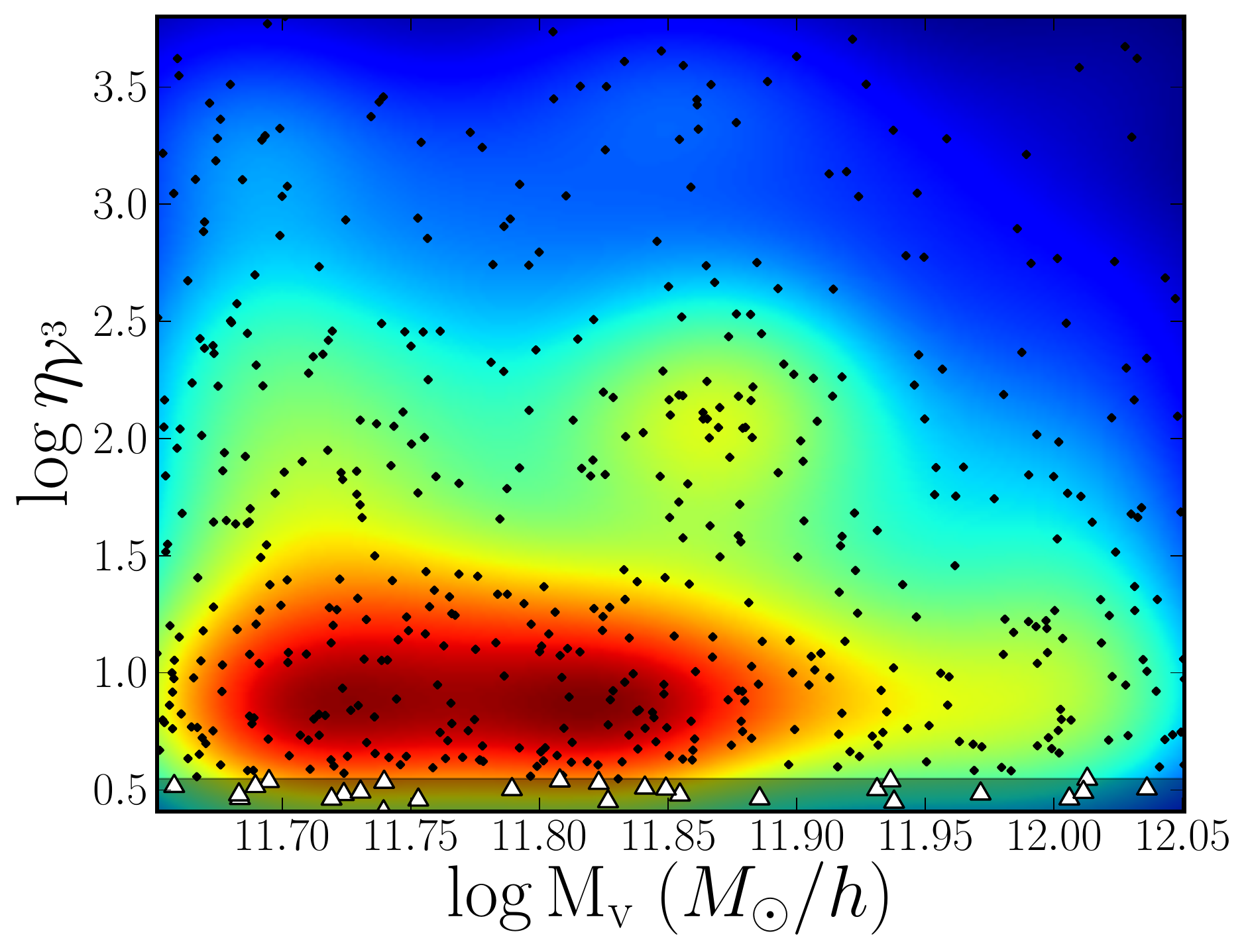}
\includegraphics[width=0.32\textwidth]{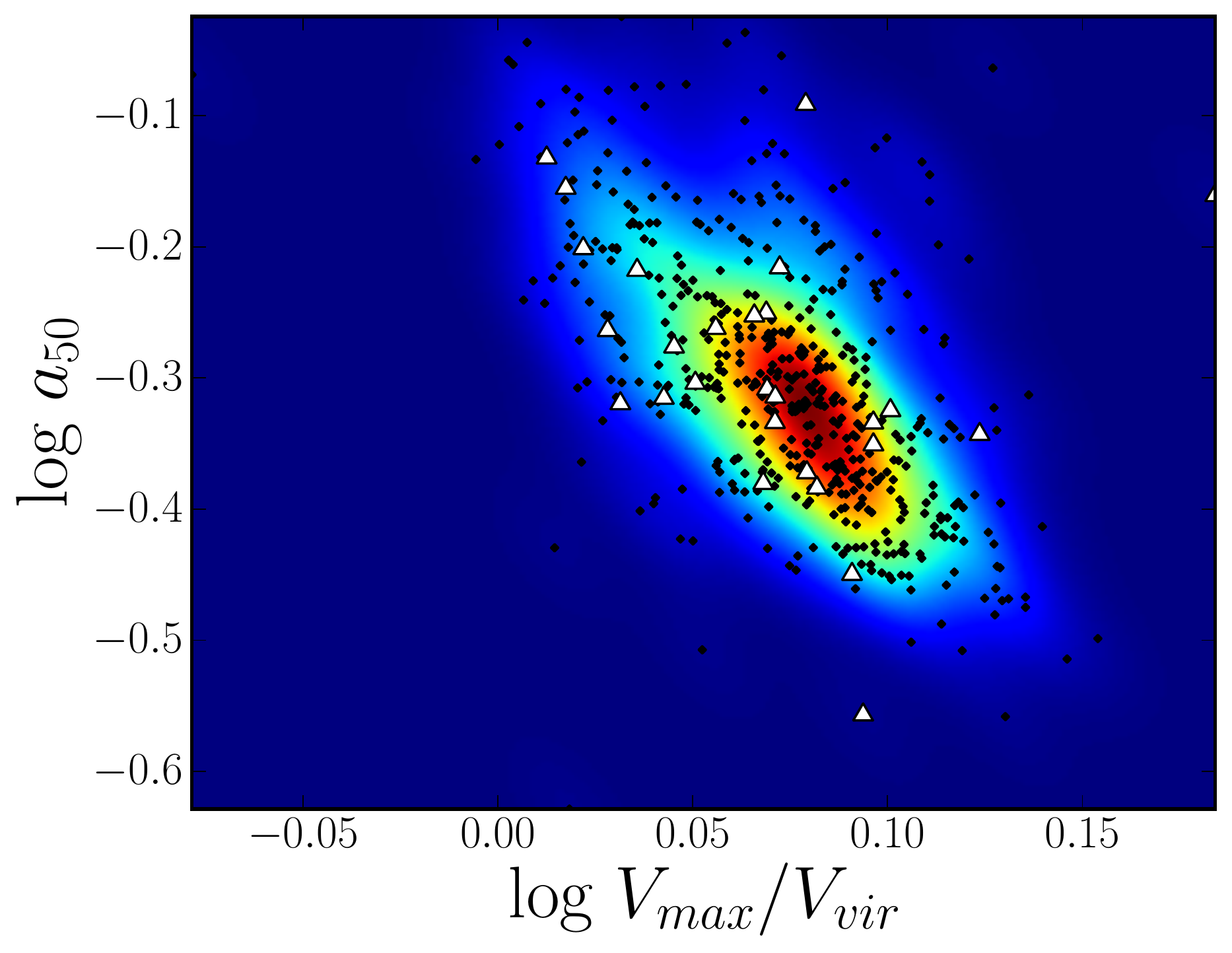}
\includegraphics[width=0.32\textwidth]{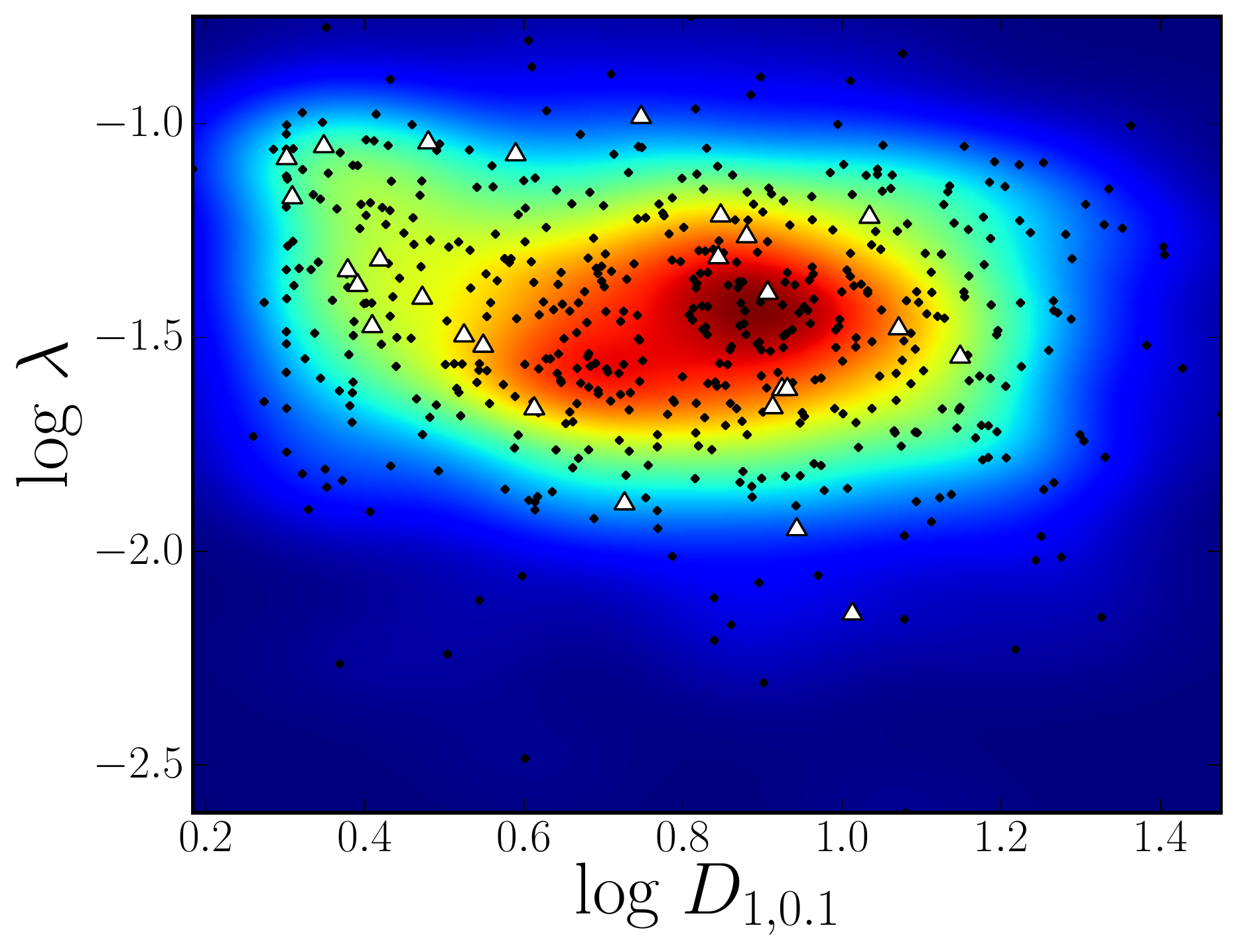}
\end{center}
\caption{Sampling halo properties with small \lagvol{} systems. In the left panel, we plot the inefficiency parameter, $\ineff$, versus the virial mass of all Milky Way size halos extracted from the L50n512 simulation. ``Small-volume" halos, i.e. those with an inefficiency, $\ineff_{\gvol^{3}}$, in the lowest $5\%$ of systems (marked by the grey band) are plotted as white triangles in all plots.  The middle panel shows a measure of halo formation time ($a_{50}$) versus concentration ($V_{max}/\vvir$); the small-volume halos exhibit no bias in these halo properties. The right panel shows the spin parameter of the halo versus an environmental parameter, $D_{1,0.1}$. Again, small-volume halos are unbiased to halo spin, but they are more likely to reside in lower density environments. In all plots, the color maps stand for the probability density function for all of the halos in the box, with dark red marking the peak of the distributions and dark blue the lowest value.}
\label{fig:halopropfixedmass}
\end{figure*}

We have just established that \lagvol{} is uncorrelated with internal halo properties.  There is some correlation with environment, but the scatter about this correlation is large.  The implication  is that the most efficient targets for zoom simulations are those with the smallest normalized \lagvol{}s --- such halos are the cheapest to run and sample the parameter space of dark matter halos in a largely unbiased way.  This point is illustrated in Figure~\ref{fig:halopropfixedmass}. The left panel shows the inefficiency for the cuboid \lagvol{}, $\ineff_{\gvol}$, as a function of virial mass, with white triangles indicating halos with $\gvol$ in the smallest $5\%$ of the distribution.  The two other panels illustrate that this subsample of small-$\gvol$
halos is statistically indistinguishable from the full sample, with  environment the lone exception.   The small-volume subsample is clearly biased towards low density environments, as expected, though some small-volume halos do exist in high density environments.

By selecting systems with small \lagvol{}  for resimulation, one can dramatically reduce the computational requirements.  The \lagvol{} of an average halo is typically $\sim 8-10$ larger than those in the $5\%$ tail, and this directly translates to unnecessary computational complexity.  Provided that the environment is not an important variable to the problem of interest, small-volume halos are ideal candidates for resimulation.  However, for problems that are closely tied to the environment, one must consider more elaborate methods to build a fair sample of halos while still considering the \lagvol{}, in order to reduce the cost of the total sample. In Table~\ref{tab:fitslvperc}, we show the fits for equation~\ref{eq:lvmass} using just the $5\%$ tail of halos with lowest \lagvol{}s of L25n512, L50n512, L650n512 and L900n512  simulations. By comparing these results with the fits presented in Table~\ref{tab:fitslv}, one can immediately see how much time and memory could be saved if the approach suggested here is taken into account.

\begin{table*}
\begin{minipage}{180mm}
\caption{Fits for equation~\ref{eq:lvmass} using just halos with the 5\% percentile lowest \lagvol{}s of L25n512, L50n512, L650n512 and L900n512  simulations.}
\label{tab:fitslvperc}
\begin{center}
\begin{tabularx}{\textwidth}{*{13}{>{\centering\arraybackslash}X}}
    \hline 
    & \multicolumn{3}{c}{Cuboid \lagvol{}} & \multicolumn{3}{c}{Minimum cuboid \lagvol{}} & \multicolumn{3}{c}{Minimum ellipsoid \lagvol{}} & \multicolumn{3}{c}{Convex hull \lagvol{}}\\
    & \multicolumn{3}{c}{($\gvol$)}& \multicolumn{3}{c}{($\grvol$)} & \multicolumn{3}{c}{($\evol$)} & \multicolumn{3}{c}{($\chvol$)}\\
    \cmidrule(lr){2-4} \cmidrule(lr){5-7} \cmidrule(lr){8-10} \cmidrule(lr){11-13}
    $\rtb$ & $1$ & $3$ & $5$ & $1$ & $3$ & $5$ & $1$ & $3$ & $5$ & $1$ & $3$ & $5$\\ 
    \hline
    $\alpha$ & -0.03 & -0.02 & -0.01  & -0.03 & -0.02& -0.03  & -0.02 & -0.01& 0.00  & -0.01& -0.01& 0.01 \\
    $A$ &$10^{2.87}$ & $10^{2.99}$ & $10^{3.13}$  & $10^{2.84}$ & $10^{2.96}$ & $10^{3.08}$ & $10^{2.76}$ & $10^{2.87}$ & $10^{2.99}$ & $10^{2.55}$ & $10^{2.67}$& $10^{2.79}$\\
    $B$ & 2.06& 2.71 & 3.69 & 1.91 & 2.50 & 3.32  & 1.59 & 2.03 & 2.67  & 1.08 & 1.28  & 1.69 \\
    \hline
  \end{tabularx}
\end{center}
\end{minipage}
\end{table*}

\subsection{Are There Pathological Halos to Avoid?}
\label{ssec:robhalo}

\begin{figure} 
\begin{center}
\includegraphics[width=0.47\textwidth]{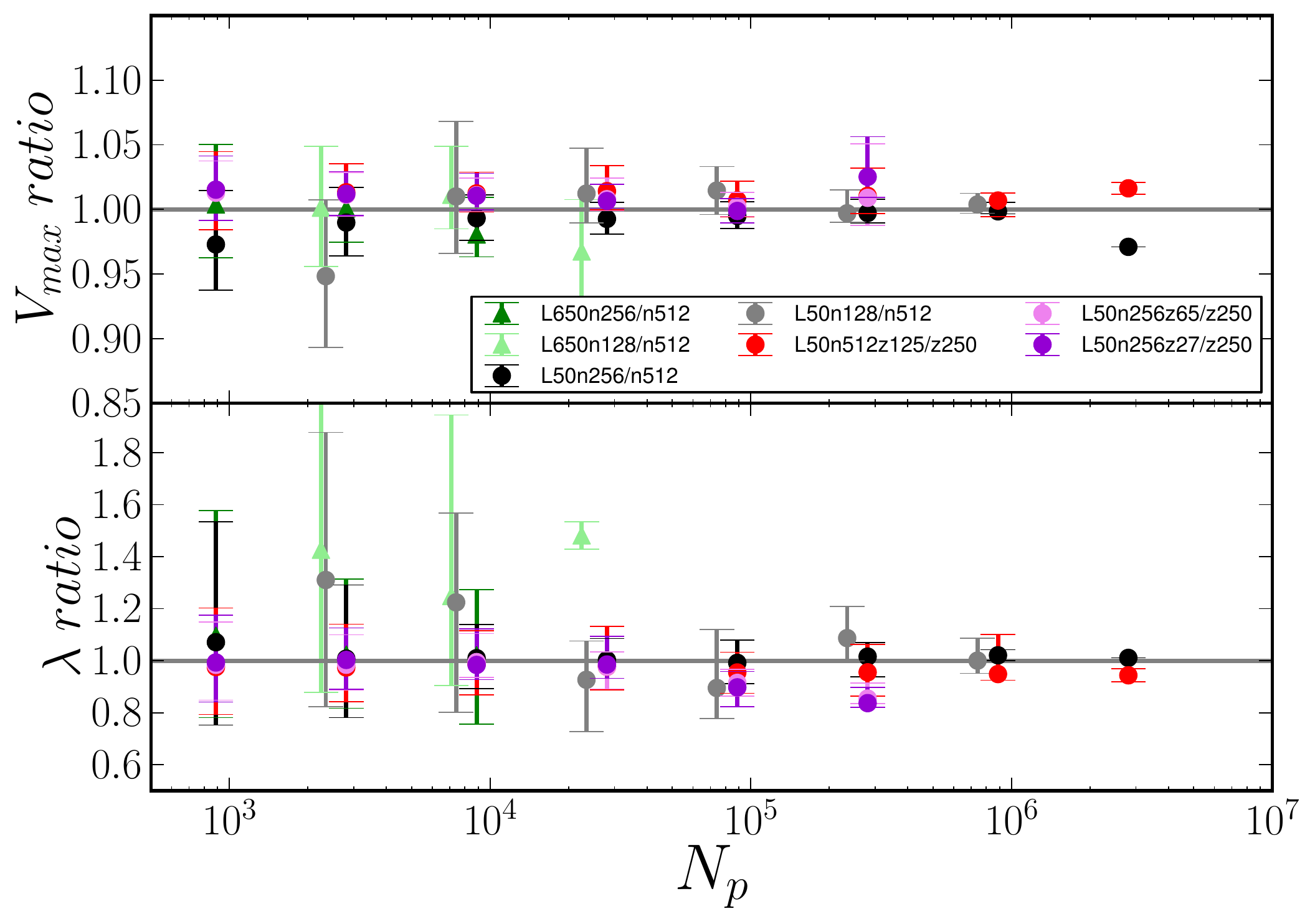}
\end{center}
\caption{Scatter in the properties of identical halos between runs using different resolution or initial redshift.
Values are binned by the number of particles, $\npart$, within $\rvir$. When comparing two different resolution runs, the high resolution $\npart$ was used. Symbols are placed at the median value and error bars indicate the $25\%$-$75\%$ spread. The upper panel shows the  change in the maximum circular velocity and lower panel the spin of a halo. We compare halos between simulations varying resolution ($512^3,256^3$ and $128^3$) for full box $L_{box}=650 \hmpc$  (green and light green triangles) and $50 \hmpc$ (black, grey circles) runs. In this case only halos that differ by less than 5\% in mass and position between runs are shown. 
We also plot differences between the halo properties of full-box simulations initialized with different initial redshifts, $\zini$. Red circles show the differences for halos between two runs with $L_{box}=50 hmpc$ and $\zini=125$ and $250$. Violet and dark violet circles show differences between a $256^{3}$,$L_{box}=50 \hmpc$ run using $\zini=27,125,250$.} 
\label{fig:haloproprob}
\end{figure}

In order to study the reliability of zoom-in simulations, specifically the effects of contamination on dark matter halos, we first needed to understand the stability of halo properties in full-box simulations.  Such properties could, in principle, be sensitive to the resolution ($\npart$), the initial redshift, or just the specific transfer function used.  We therefore re-ran several of our full-box simulations, varying these parameters. We obtain a reasonable estimate for the resultant scatter by comparing the properties of identical halos in these different runs, which we use to determine the importance of contamination, \lagvol{} definitions, and sizes of buffer zones in the multimass simulations.  Figure~\ref{fig:haloproprob}, which we discuss throughout this section, illustrates a few of these tests.

Trajectories within virialized systems are chaotic.  Small differences existing at any time grow quickly. These differences originate in the dynamical instability of particle trajectories in high-density regions \citep[e.g.,][]{Knebe:2000,Valluri:2007}; in particular, as was first noted by \citet{Miller:1964}, the N-body problem is chaotic in the sense that the trajectory of the 6N-dimensional phase-space coordinate of the system exhibits exponential sensitivity under small changes in the initial conditions. This exponential instability, referred to in the literature as the Miller instability or  ``minichaos'', has been investigated extensively in several studies over the past three decades \citep[see][for a review]{Merritt:2005}, which have shown that this chaotic behavior is not reduced  by increasing the number of particles \citep{Kandrup:1991,Goodman:1993}.  Therefore changes in the initial conditions, even for a fixed resolution (e.g. using a different $\zini$), will cause the particles to experience slightly different trajectories, which produces an artificial scatter in the halo properties. 

Most halo properties exhibit a relatively small scatter as the resolution and $\zini$ vary, with median values in the ratio of structural quantities of approximately unity, as illustrated in  the upper panel of Figure~\ref{fig:haloproprob}. However, properties that are strongly affected by substructure, and therefore the specific positions of the particles, such as the spin or the halo shape (lower panel of Figure~\ref{fig:haloproprob}), while still convergent, show a significantly larger scatter about the median.  The scatter in these properties can be reduced significantly if particles bound to substructures are eliminated from the calculation \citep[e.g.][]{VeraCiro:2011}.

For full-box halo samples, one can safely conclude that halo properties do not vary in any systematic sense. Nevertheless, at least two other interesting questions exist for multimass simulations.  First, do the variations between simulations correlate with any specific halo property?  That is, are there halos that are most likely to have a large scatter between runs be identified beforehand by any $z=0$ property?  Second, do those halos that exhibit a large dispersion between two runs also exhibit substantial scatter when simulated with a third $\zini$ or at a different resolution?  Specifically, do certain halos tend to have a larger dispersion between their parameters?  If either is the case, one would like to identify such halos and avoid them when selecting halos for resimulation.  Our analysis of the full-box simulations presented here indicates that neither is true:  we find that no $z = 0$ property is correlated with a larger dispersion in halo parameters after a resimulation at different resolution or with a different $\zini$. We also found that when comparing halos between three different re-simulations there are no specific halos that tend to have a bigger dispersion. Our results are consistent with the hypothesis that variations are driven by the number of particles that begin with high density fluctuations \citep{Knebe:2009}.

\section{Creating Multimass Initial Conditions}
\label{sec:createICs}

\subsection{Starting Redshift and Lagrange Volume}
\label{ssec:zstart}

The initial redshift, $\zini$, is one of many parameters that must be chosen for cosmological simulations.  Usually, $\zini$ is chosen such that the resulting RMS variance of the discrete density field\footnote{$\sigma_{L}^{2}=\frac{1}{2\pi^{2}}\int^{k_{max}}_{k_{min}}P(k)k^{2}dk$ where $k_{min}=2\pi/L$ represents the fundamental mode determined by the box size, L, and, $k_{max} = \pi N^{1/3}/L$, the Nyquist frequency that additionally depends on the number of particles, N, used for the initial conditions.}, $\sigma_{L}$, is between $0.1$ and $0.2$ \citep[see][]{Knebe:2009}. Though one might expect that a higher $\zini$ is necessarily better, lower initial redshifts are computationally cheaper.  Moreover, beginning too early can introduce additional round-off errors and shot noise in the particles used to sample the primordial matter density field \citep{Lukic:2007}. If the simulation is started too late, however, non-linear evolution will be compromised, delaying the collapse of the first halos that act  as seeds for further structure formation \citep{Jenkins:2010,Reed:2013}.
 
Improvements in the techniques used to calculate perturbations have reduced errors in the quasi-linear regime and allowed for lower values of $\zini$ \citep[see, for example][]{Jenkins:2010}; some groups \citep[e.g.][]{Lukic:2007,Prunet:2008,Knebe:2009,Jenkins:2010,Reed:2013} have worked to quantify the effect of $\zini$ on the final results of cosmological simulations. 
In fact, the specific starting redshift of a given cosmological simulation depends upon the specific technique used to calculate the perturbations, the specific scientific problem at hand, and the redshift of the first analyzed output \citep{Reed:2013}.  Typically, multimass simulations are additionally constrained to use the same $\zini$ as that of the full-box simulation from which the target halo was selected, as the calculated \lagvol{} is only exactly valid at that redshift.


Our approach for selecting $\zini$ is to compute the values of $\zini^{0.1}$ and $\zini^{0.2}$ using the full box at the lowest resolution (that used to calculate the \lagvol{}) and the highest resolution of relevance (corresponding to the zoom-in simulation). These two redshift ranges will overlap unless $\Delta_{res}$ is particularly large (which is not recommended). 
For example, a box size of $650 \hmpc$ with our adopted cosmology gives initial redshift values of $\zini^{0.1}=27.81$ and $\zini^{0.2}=13.40$ for $N_{p}=512^3$ run, and the same simulation with an effective resolution of $2048^3$ corresponds to $\zini^{0.1}=46.10 - \zini^{0.2}=22.55$. Therefore an initial redshift between $27.81$ and $22.55$ would be appropriate for both simulations, assuming we are interested in $z=0$ results.

\begin{figure} 
\begin{center}
\includegraphics[width=0.47\textwidth]{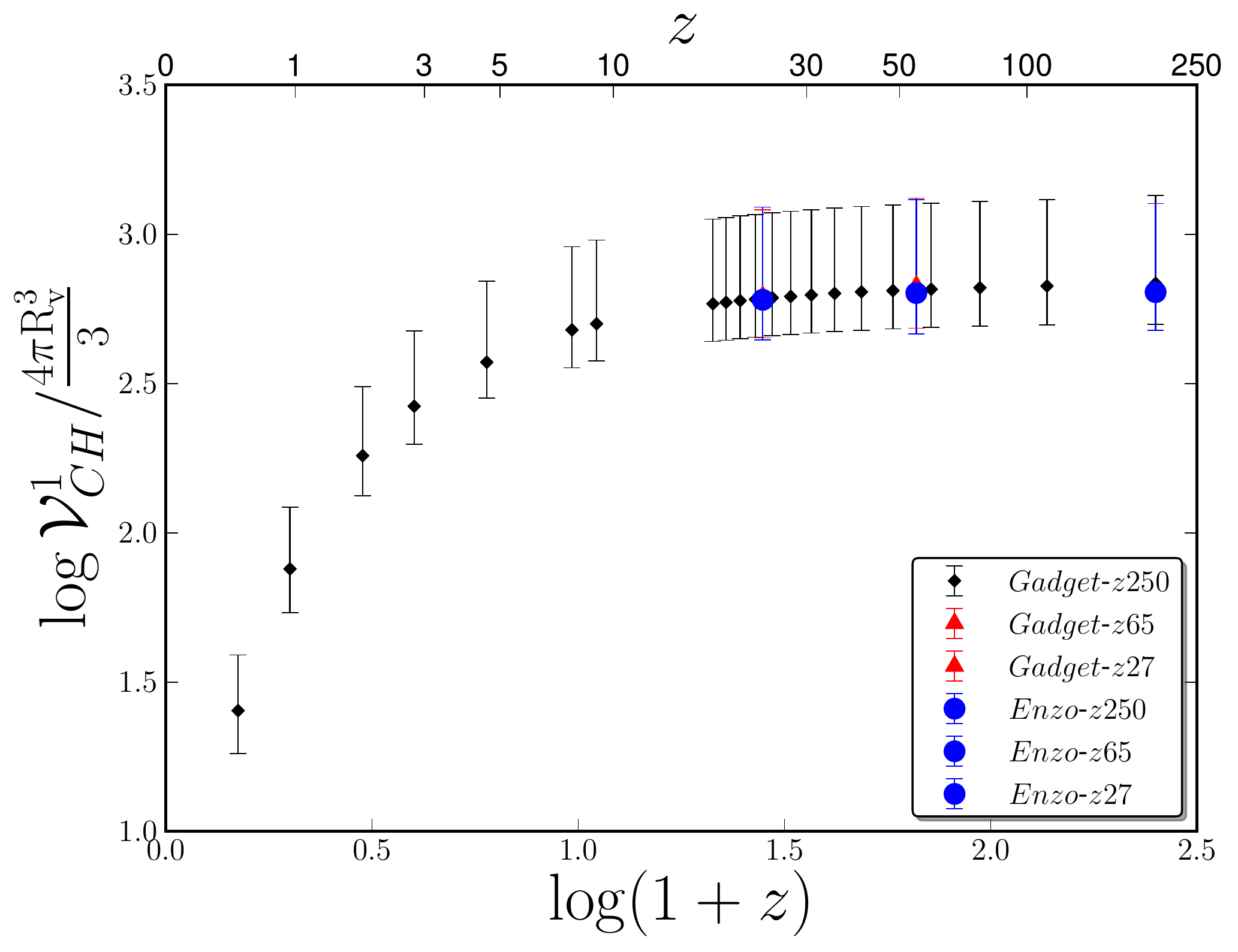}
\end{center}
\caption{Redshift evolution of the \lagvol{}. Plotted is the normalized convex hull \lagvol{} for halos at $z=0$ in the  L50n256 simulation versus the redshift at which the volume is calculated.  Black diamonds indicate halos from L50n256  \gadget{} run with $\zini = 250$, red triangles signify L50n256 \gadget{} runs with $\zini = 125$ and $\zini = 27$. 
The blue circles are equivalent results obtained from \enzo{} (which agree perfectly with results from \gadget).  The \lagvol{} appears independent of the initial redshift of the simulation, and remains roughly constant until the turn-around redshift, the exact value of which varies with halo mass. In this case the turn around point indicates halos with mass $\mvir\sim3.55\times10^{11} \hmsun $ which comprise the majority of the halo population in the L50n256 simulations.}
\label{fig:lagvolzevol}
\end{figure}

We have re-run several full-box simulations with different initial redshift to ensure that the different results and trends found in this work do not depend on the specific $\zini$ chosen. In Figure~\ref{fig:lagvolzevol}, we plot the evolution of the \lagvol{} of a L50n256 simulation for halos at $z=0$ (black points) along with the $\zini$ \lagvol{} computed for two other identical L50n256 simulations but with a lower $\zini$ (red triangles).
The calculated \lagvol{} is largely independent of the initial redshift and, in fact, $\gvol$ evolves very slowly for redshifts greater than the (mass dependent) turn-around redshift; therefore, all of our results concerning the \lagvol{} are valid for any $\zini$ greater than the turn-around redshift (therefore including $\zini=\infty$). In this Figure we also plot the $\zini$ \lagvol{} computed for three identical L50n256 simulations runs using \enzo{} (blue circles) which show a very good agreement with \gadget{} results.

\subsection{Lower resolution regions}
\label{ssec:buffer}
Initial conditions for multimass simulations typically surround the high resolution region with shells, or buffer volumes, of progressively lower resolution particles, with the number of shells varying from $0$ to $\Delta_{res}-1$.  Again, many groups have developed a methodology based on trial and error, but our literature search revealed no quantitative analysis.  We explore how the properties of these buffer volumes, specifically the size, number, and steps in resolution between them, affect uncontaminated halos at $z = 0$.

Our results indicate that using a step in resolution larger than one between successive shells, i.e. a factor of $8^2$ or greater increase in particle mass, strongly alters the properties of the halo at $z = 0$.  It is unclear whether the variance in the high resolution region is due to the mixing of dramatically different particle masses, which leads to poorly resolved structures in the buffer volumes, or because the low mass particles near the edge of the high resolution volume have their trajectories too strongly affected by the high mass particles nearby, but the effects also propagate into uncontaminated regions. Though larger jumps in resolution in the lower resolution regions may work in some cases, our tests indicated that such a technique is generally problematic.  Therefore, we recommend using a buffer zone for each step in resolution (factor eight in mass) between the largest and smallest particle masses.

Similarly, the host halo will not be properly converged if the buffer volumes are too small; however, increasing the size beyond this minimum value does not effect the final properties in any way but increases the CPU and memory cost of the simulation. As is the default in \music{}, we created multimass initial conditions such that the width of a given level, $w_i$, sets the width of next lower resolution grid level, $w_{i-1}$.  In units of the grid size of the highest resolution grid in the simulation (i.e. the \lagvol{}), 
\[
w_{i-1} = w_i/2 + c,
\]
where $c$ is a constant added to guarantee that each shell has a minimum width; that is, the second highest resolution region is half as wide ($+c$) as the \lagvol{} in the initial conditions.  Test multimass simulations show that variations in $c$ do not alter the final halo, provided that $c \geq 12$.  For users of \music{}, we note that this corresponds to \texttt{padding $\geq$ 6}, as $c = 2\times$\texttt{padding}.

Finally, using the recipe described above, we varied the lowest resolution level that comprises the majority of the multimass box, which fixes the number of buffer zones. As expected, we found that there is generally a minimum for this resolution: it must be high enough to capture large-scale tidal forces relevant to the targeted halo. The multimass simulations studied in this work indicate that the host halo is converged provided the lowest effective resolution is at least $128^3$. Again, setting this resolution higher results in unnecessary, and significant, memory and CPU cost.

\subsection{Other Definitions of Lagrange Volume}
\label{ssec:zoomchlagvol}


We also present results of multimass simulations using different \lagvol{} definitions. Though we have not run as many simulations for these cases, there are a number of interesting results. We have initialized multimass simulations using the convex hull \lagvol{}s instead of the cuboid by creating full-box initial conditions at all required resolutions, selecting from the highest resolution box all particles that lie within $\chvol$, then filling the remainder of the volume with particles from the low resolution initial conditions. We ran twelve \gadget{} simulations of halos with mass $\mvir \simeq 7.1\times10^{11} \hmsun$.  These are not listed in Table~\ref{tab:zooms}. For a fixed $\rtb$, this technique increases the likelihood of contamination compared with the cuboid approach. It therefore  requires larger values of $\rtb$ to avoid contamination. However, even taking this into account, we found that using this technique, one may typically resolve uncontaminated halos with $40-50\%$ fewer particles than with the cuboid \lagvol{}. To ensure uncontaminated halos using this technique (in parallel with Equation~\ref{eq:lvol}), we recommend a convex hull \lagvol{} defined by
\begin{equation}
\rtb = (1.5*\Delta_{res}+7)\times \rvir.
\label{eq:lvolch}
\end{equation}

\texttt{MUSIC} has recently added support to generate initial conditions using also the minimum ellipsoid \lagvol{} ($\evol$). Running a full set of simulations using this \lagvol{} definition is beyond the scope of this paper. However  the convex hull results (Eq.~\ref{eq:lvolch}) will ensure uncontaminated halos for this \lagvol{} definition as well.

\subsection{Eulerian Codes}
\label{ssec:zoomcodes}

We have also run full box and multimass simulations using \enzo. We already showed that, for the same initial conditions, \lagvol{}s obtained using \enzo{} or \gadget{} are very similar. In the case of multimass simulations, the similar effects of low resolution contamination were also found with \enzo; that is, for a fixed \lagvol{}, low resolution particles get closer to the center of the halo as the level of zoom increases. Therefore the difference in resolution between that used to calculate the \lagvol{} and that used in the high resolution region of the multimass simulation has to be taken into account. We also found that due to the adaptive mesh refinement algorithm (AMR) in Eulerian codes, it is recommendable to include the region in which the halo is located along redshift
when calculating the \lagvol{}. We studied this effect and found that the distance between the center of mass at $\zini$ and $z=0$ only shows a weak correlation with the environmental parameter, $N^{3}_{neigh}$. Therefore, this parameter can also be used to pick halos for zoom-in simulations without significantly biasing the final sample. Finally, in the case of \enzo{} we find much more scatter between similar zoom-in simulations and therefore do not want to give an explicit recommendation concerning the size of the \lagvol{}. The study of a higher statistical \enzo{} sample is beyond the scope of this paper.

\section{Conclusion and Discussion}
\label{sec:conc}
We have run a large number of cosmological full-box and zoom-in simulations of various sizes, particle number, and initial redshifts.  By analyzing the properties of the halos in these simulations, we studied the methodology behind the multimass technique and have arrived at a number of conclusions and recommendations regarding the \lagvol{}, contamination, and how to best perform zoom simulations.

In the first part of this paper we presented a detailed study of the high-redshift \lagvol{} associated with the makeup of $z=0$ halos, including a variety of definitions for calculating the  \lagvol{} itself.  Importantly, we have found that the \lagvol{} of any given halo converges very slowly as the mass resolution of the simulation increases, and that there can be catastrophic errors if fewer than $\sim 500$ particles are used.   Though halo virial masses are well-converged, \lagvol{} regions can be quite irregular, with overall volumes governed by the furthest few particles.   Extrapolating the results of Figure 4, we find that approximately $\sim 10^6$ particles are required in a {\em low resolution} halo in order to have a confidently converged \lagvol{}.  This result is related to a particularly striking finding (Figure 6): that the chance for low-resolution particle contamination increase steadily with the level of zoom.
Equation~\ref{eq:lvol} provides a formula for choosing the \lagvol{} conservatively to avoid halo contamination based on the level of refinement between the parent simulation and the ultimate resimulation volume.  This formula was derived using $\sim130$ zoom-in simulations that focused on objects ranging from dwarf galaxy halos to clusters, and is applicable for Lagrangian codes.



In studying the general properties of dark-matter only halos with 1-2\% contamination by mass, we do not find significant variation in the fundamental halo properties (formation time, concentration, shape, etc.). However, we do find important divergences between the halo properties of contaminated and non-contaminated halos in adiabatic gas multimass runs, such as spurious baryonic structure formation, and therefore strongly recommend the usage of \lagvol{}s that guarantee uncontaminated halos for all hydrodynamic simulations.

We have also performed a detailed statistical analysis between the varying \lagvol{} definitions and other typical halo properties using a sample of halos spanning eight orders of magnitude in mass from a set of full-box simulations.  After normalizing the \lagvol{} by the virial volume of each halo, we find an anticorrelation between \lagvol{} and virial mass. Smaller halos therefore have proportionally larger \lagvol{}s than more massive ones. Thus, cluster zoom-in simulations are cheaper than those focused on dwarf galaxy halos, for a fixed number of particles within the halo. However, at fixed halo mass, the \lagvol{} does not correlate with any internal halo property we have considered.  We find a mild correlation with large-scale environment, such that the most clustered halos have larger \lagvol{}s.  

Based on our results we provide provide the following recommendations for performing multi mass zoom-in simulations:

\begin{itemize}
 \item Run the full-box simulation from which the zoom-in halos will be selected at the highest resolution possible compared to the ultimate zoom-in simulation. This will allow for more accurate values of the \lagvol{}s to build initial conditions and decrease the chances for contamination.  
 \item Select the initial redshift, $\zini$, for the full-box simulation based on the highest resolution zoom-in simulation planned.  Though the \lagvol{} evolves slowly at early times, a calculated \lagvol{} is only exactly correct at that redshift.
 \item Do not pick halos for zoom-in simulations with fewer than 500 particles in the full-box run -- the \lagvol{} will dramatically increase with increasing resolution for such poorly resolved halos.
 \item Avoid contamination in the final halo by accounting for the difference in resolution between the full-box and the zoom-in simulation when selecting the \lagvol{} (see Equations~\ref{eq:lvol}~and~\ref{eq:lvolch}).
 \item For a given halo mass, select those halos with the smallest \lagvol{}s for zoom-in simulations, in order to minimize CPU and memory cost.
 \item Create a buffer volume at each resolution step between the high resolution region and the lowest resolution that comprises the majority of the box; however, the size of these volumes and the lowest effective resolution should both be
minimized to again eliminate unnecessary CPU and memory usage. The simulations analyzed in this work indicate that using $128^{3}$ for the latter effectively captures the large-scale tidal forces; we elaborate on the minimum width in Section~\ref{ssec:buffer}, but note for users of \texttt{MUSIC} that our results require \texttt{padding} $\geq 6$. 
 \item In the case of the most expensive multimass hydrodynamical simulations it is highly advisable to first run the collisionless version with the same \lagvol{} to guarantee non-contaminated halos. 

\end{itemize}


\section*{Acknowledgments}
We want to thank the referee for all comments and suggestions that helped improve this paper.
This work used computational resources granted by NASA Advanced Supercomputing (NAS) Division, Nasa Center for Climate Simulation, Teragrid and by the Extreme Science and Engineering Discovery Environment (XSEDE), which is supported by National Science Foundation grant number OCI-1053575. 
We also made use of resources at the City University of New York High Performance Computing Center which is supported, in part,  under National Science Foundation Grants CNS-0958379 and CNS-0855217 and the GreenPlanet cluster at UCI.
JO, SGK, JSB, and MR were supported by NSF grant AST-1009999 and NASA grant NNX09AG01G. 
JO also thanks the financial support of the Fulbright/MICINN Program.
OH acknowledges support from the Swiss National Science Foundation (SNSF) through the Ambizione fellowship.

\bibliography{howtozoom}
\bibliographystyle{mn2e}

\end{document}